\documentclass[preprint,12pt]{elsarticle}

\usepackage[a4paper, left=2.5cm, right=2.5cm, top=3.3cm,  bottom=3.3cm]{geometry}

\biboptions{sort&compress}

\usepackage{amsmath,amssymb,bm}
\usepackage{bbm}
\usepackage{graphicx}
\usepackage{float}
\usepackage{subcaption}

\usepackage{tabularx}
\usepackage{booktabs}
\usepackage{multirow}
\usepackage{array}
\usepackage{makecell,tabularx}
\usepackage{adjustbox}

\usepackage{algorithm}
\usepackage{algpseudocode}

\usepackage{stackengine}

\usepackage{pgfplots}
\pgfplotsset{compat=1.18}
\usepackage{xcolor}

\usepackage{tikz}
\usetikzlibrary{positioning, fit, backgrounds, arrows.meta, shapes, calc}
\usepackage{forest}

\usepackage{enumitem}
\usepackage{calc}

\usepackage{hyperref}
\usepackage{cleveref}

\newcommand{\mathbbmI}{\mathbb{I}}

\newcolumntype{C}[1]{>{\centering\arraybackslash}p{#1}}
\newcolumntype{L}[1]{>{\raggedright\arraybackslash}p{#1}}

\journal{}

\begin{document}

\begin{frontmatter}

\title{Regime Discovery and Intra-Regime Return Dynamics in Global Equity Markets}

\author[aff1]{S.~R.~Luwang\corref{cor1}}
\ead{salamrabindrajit@gmail.com}

\author[aff1]{B.~N.~Sharma}
\ead{bnsharma09@yahoo.com}

\author[aff1]{K.~Mukhia}
\ead{kundanmukhia07@gmail.com}

\author[aff1]{Md.~Nurujjaman}
\ead{md.nurujjaman@nitsikkim.ac.in}

\author[aff2]{Anish~Rai}
\ead{anishrai412@gmail.com}

\author[aff3]{Filippo~Petroni}
\ead{fpetroni@luiss.it}

\author[aff4,aff5]{Luis~E.~C.~Rocha}
\ead{luis.rocha@ugent.be}

\cortext[cor1]{Corresponding author.}

\affiliation[aff1]{organization={Department of Physics, National Institute of Technology Sikkim},
            postcode={737139},
            country={India}}

\affiliation[aff2]{organization={Chennai Mathematical Institute},
            postcode={603103},
            country={India}}

\affiliation[aff3]{organization={Department of Economics, University G. d’Annunzio of Chieti-Pescara},
            postcode={65127},
            country={Italy}}

\affiliation[aff4]{organization={Department of Economics, Ghent University},
            postcode={9000},
            country={Belgium}}

\affiliation[aff5]{organization={Department of Physics and Astronomy, Ghent University},
            postcode={9000},
            country={Belgium}}

\begin{abstract}
Financial markets alternate between tranquil periods and episodes of stress, and return dynamics can change substantially across these regimes. We study regime-dependent dynamics in developed and developing equity indices using a data-driven Hilbert--Huang-based regime identification and profiling pipeline, followed by variable-length Markov modeling of categorized returns. Market regimes are identified using an Empirical Mode Decomposition-based Hilbert--Huang Transform, where instantaneous energy from the Hilbert spectrum separates Normal, High, and Extreme regimes. We then profile each regime using Holo--Hilbert Spectral Analysis, which jointly resolves carrier frequencies, amplitude-modulation frequencies, and amplitude-modulation energy (AME). AME, interpreted as volatility intensity, declines monotonically from Extreme to High to Normal regimes. This decline is markedly sharper in developed markets, while developing markets retain higher baseline volatility intensity even in Normal regimes. Building on these regime-specific volatility signatures, we discretize daily returns into five quintile states $\mathtt{R}_1$ to $\mathtt{R}_5$ and estimate Variable-Length Markov Chains via context trees within each regime. Unconditional state probabilities show tail states dominate in Extreme regimes and recede as regimes stabilize, alongside persistent downside asymmetry. Entropy peaks in High regimes, indicating maximum unpredictability during moderate-volatility periods. Conditional transition dynamics, evaluated over contexts of length up to three days from the context-tree estimates, indicate that developed markets normalize more effectively as stress subsides, whereas developing markets retain residual tail dependence and downside persistence even in Normal regimes, consistent with a coexistence of continuation and burst-like shifts. Overall, market maturity shapes both the pace of stabilization and the persistence of tail dependence, supporting tighter risk controls not only during crises but also during periods classified as stable.
\end{abstract}


\begin{keyword}
Hilbert-Huang Transform \sep Holo-Hilbert Spectral Analysis \sep volatility intensity \sep market regimes \sep variable-length Markov chains \sep context trees 
\end{keyword}

\end{frontmatter}

\section{Introduction}
\label{sec:Intro}

Complex systems such as the stock market rarely move along one smooth and simple path. They typically alternate between distinct regimes -- periods of relatively stable conditions and episodes of rapid growth or sudden collapse~\cite{hamilton1990analysis,ang2012regime}. During stable regimes in the stock market, volatility remains low, asset returns are moderate, and liquidity is abundant. On the other hand, turbulent regimes are marked by sharp price swings, widening bid-ask spreads, and heightened probabilities of extreme outcomes -- losses or gains~\cite{engle2004risk,longin1996asymptotic}. These unique statistical profiles influence how returns transition between categories -- such as extreme losses, minor fluctuations, or substantial gains, within a regime~\cite{rey2014detection,hamilton1989new,bensaida2015frequency}. Consequently, investment strategies and risk management calibrated for one regime may perform poorly or become risky when market conditions shift~\cite{ang2012regime,kritzman2012regime,nystrup2015regime}. These considerations highlight the importance of identifying market regimes and analyzing return dynamics within each regime to manage risks better and improve portfolio decisions, which are key steps for navigating the uncertain nature of stock markets.

A broad literature has examined stock market return dynamics using diverse methodological frameworks~\cite{ang2007stock,flannery2002macroeconomic,marquering2004economic,avramov2006predicting,rabindrajit2024high}. Markov-based models have shown to be particularly effective in capturing regime-dependent behavior. Existing studies have used Markov regime-switching specifications to model shifts in conditional variance~\cite{chang2009macroeconomic}, Markov-chain formulations to test persistence and deviations from the random-walk hypothesis~\cite{mcqueen1991stock}, and forecasting frameworks based on regular and absorbing Markov chains~\cite{huang2017applying}. Extensions that incorporate memory effects, including semi-Markov and indexed Markov formulations, further enrich this line of work by allowing history-dependent and duration-dependent dynamics, with particular relevance for high-frequency financial data~\cite{d2019change,d2018copula,d2012weighted,d2011semi}. Related applications also examine asymmetric transmission channels such as the impact of oil price shocks on stock returns~\cite{reboredo2010nonlinear}. Despite these advances, many Markov-based approaches still impose a fixed conditioning structure on transitions, including fixed order, prescribed indexing, or parametric kernels, which can limit their ability to capture complex and adaptive temporal dependencies in return dynamics. Variable-length Markov chain (VLMC), which allow the effective memory length to vary according to statistically significant contexts, offers a flexible alternative. However, to the best of our knowledge, VLMCs have not yet been employed to analyze transitions between return categories such as extreme losses, minor fluctuations, and substantial gains across distinct market regimes.

To analyze regime-specific transition dynamics between return categories, market regimes must first be reliably identified and differentiated. In this paper, we employ the Empirical mode decomposition-based Hilbert-Huang Transform (HHT) framework for regime identification~\cite{rai2023detection}. While index return series are often weakly stationary~\cite{cont2001empirical}, their volatility dynamics are highly time-varying and exhibit intermittent bursts, particularly during periods of market stress~\cite{mandelbrot1963variation,schwert1989does}. Abrupt changes in fluctuation intensity and short-lived episodes of elevated activity are therefore central features of financial returns rather than rare anomalies~\cite{lux2000volatility,engle2007good}. Standard volatility measures, including rolling standard deviations or GARCH-type conditional variances, provide smoothed and model-dependent estimates of risk intensity and are primarily designed to capture gradual volatility clustering~\cite{andersen1998answering,mikosch2004nonstationarities,gatheral2011volatility}. In contrast, the HHT framework offers an adaptive, data-driven time-frequency representation that is sensitive to rapid changes in oscillatory amplitude. By decomposing the return series into intrinsic mode functions and applying the Hilbert Transform, HHT yields instantaneous amplitudes, frequencies, and energies (squared instantaneous amplitudes)~\cite{rai2023detection,mahata2021characteristics}. The resulting instantaneous energy captures the time-localized concentration of oscillatory activity and serves as a natural proxy for the intensity of market fluctuations. Using this instantaneous energy, we identify three market regimes -- Normal, High, and Extreme.

Identifying regimes from instantaneous energy provides an initial segmentation of market states. However, this does not explain how the regimes differ internally. We therefore examine whether the identified regimes exhibit distinct volatility structure by quantifying regime-specific volatility intensity, defined as the magnitude of volatility fluctuations over time, using Holo-Hilbert Spectral Analysis (HHSA)~\cite{huang2016holo}. HHSA extends the HHT framework by capturing amplitude-modulation dynamics across intrinsic scales. In this representation, the carrier frequency corresponds to dominant oscillatory time scales of price dynamics, while the amplitude-modulation frequency describes temporal variations in oscillation strength, reflecting volatility clustering and cross-scale feedback. The squared amplitude of the modulation component, integrated over time, yields the amplitude-modulation energy. We interpret this energy as a scale-resolved proxy for volatility intensity, enabling differentiation of regimes through their volatility signatures and providing empirical support that the regimes display distinct internal dynamics. This, in turn, strengthens the motivation for subsequently investigating intra-regime return dynamics using VLMC. While HHT has been applied in financial contexts~\cite{rai2023detection,mahata2021characteristics}, the use of HHSA for regime profiling and volatility quantification remains largely unexplored in finance~\cite{chang2022evaluating,zheng2023multiscale,lee2022full,ying2024order}. 

Following regime identification and regime profiling, we analyze regime-specific return dynamics across developed and developing markets. We use daily return data from January 2000 to April 2025 for twenty stock market indices, comprising ten developed and ten developing economies. Returns are discretized into five categories: extreme loss ($\mathtt{R}_1$), mild loss ($\mathtt{R}_2$), no change ($\mathtt{R}_3$), mild gain ($\mathtt{R}_4$), and extreme gain ($\mathtt{R}_5$). VLMC is then employed to model transitions between these five states within each regime. We compare unconditional state probabilities using tail ratios and Shannon entropy, and analyze conditional transition dynamics through order-specific metrics. Self-persistence and Mean-reversion capture first-order dynamics, while Continuation, Exhaustion, Zigzag Alternation, and Burst-from-Calm capture higher-order transition behavior.

The main contributions of this paper are: First, we pioneer the application of Holo-Hilbert Spectral Analysis to finance by using amplitude-modulation energy to quantify the volatility intensity of stock market regimes. Second, we introduce Variable-length Markov chains into financial regime analysis to decode intra-regime return dynamics, and define novel metrics to quantify higher-order transition behavior. Third, we provide a comparative analysis of developed and developing markets, revealing systematic differences in regime-dependent return dynamics with implications for risk management.

The remainder of the paper is organized as follows. Section~\ref{sec:Method} presents the data \& return categorization and overall the methodological framework. Section~\ref{sec:R} presents the empirical results. In Section~\ref{sec:Disc}, we discuss the merits of our methodological framework and evaluates the robustness of the thresholds employed for regime identification. Section~\ref{sec:Conc} concludes the study.

\section{Methodology}
\label{sec:Method}

In this section, we describe the data and the methodological pipeline used in this study. We first introduce the stock-index data set and the construction of daily log returns. We then present the Empirical Mode Decomposition based Hilbert--Huang Transform framework used for regime identification via instantaneous energy, followed by Holo--Hilbert Spectral Analysis for profiling the identified regimes through their cross-frequency volatility signatures. Finally, we describe the regime-specific return-dynamics analysis, where returns are discretized into quintile-based states and modeled using Variable-Length Markov Chains estimated via context trees, together with the metrics used to summarize the inferred transition structure across regimes.

\subsection{Data}
\label{sec:DD}
We analyze daily closing prices of the standard stock market indices for twenty countries—ten developed and ten developing economies, as classified by the World Economic Situation and Prospects 2025 (United Nations) report ~\cite{undesa2025wesp,fantom2016world}. The daily data is analyzed from January 2000 to April 2025. The list of the indices can be seen from Table~\ref{tab:indices}. All of these data are downloaded from and freely available at \href{https://finance.yahoo.com/}{Yahoo Finance}.

\begin{table}[!htbp]
\centering
\footnotesize 
\setlength{\tabcolsep}{3pt}        
\renewcommand{\arraystretch}{1.1} 

\begin{tabularx}{\linewidth}{|l|>{\raggedright\arraybackslash}X|l|l|>{\raggedright\arraybackslash}X|l|}
\hline
\multicolumn{3}{|c|}{Developed economies} & \multicolumn{3}{c|}{Developing economies} \\
\hline
Country & Index & Ticker & Country & Index & Ticker \\
\hline
Australia      & S\&P/ASX 200       & AXJO      & Brazil        & Bovespa           & BVSP \\
Belgium        & BEL 20             & BFX       & Indonesia     & Jakarta Composite & JKSE \\
France         & CAC 40             & FCHI      & Argentina     & MERVAL            & MERV \\
United Kingdom & FTSE 100           & FTSE      & Mexico        & IPC               & MXX  \\
Germany        & DAX                & GDAXI     & Thailand      & SET               & SET.BK \\
Spain          & IBEX 35            & IBEX      & Singapore     & Straits Times     & STI \\
South Korea    & KOSPI              & KS11      & Saudi Arabia  & TASI              & TASI.SR \\
Japan          & Nikkei 225         & N225      & Taiwan        & TAIEX             & TWII \\
United States  & NYSE Composite     & NYA       & China         & SSE Composite     & 000001.SS \\
Switzerland    & SMI                & SSMI      & Hong Kong     & HSI               & 0388.HK \\
\hline
\end{tabularx}

\caption{Stock market indices in developed and developing economies, selected for this study.}
\label{tab:indices}
\end{table}

For each index, let \(P_t\) denote the closing price on trading day \(t\). We compute the one-day continuously compounded return as
\begin{equation}
\mathtt{r}_t = \ln\!\left(\frac{P_t}{P_{t-1}}\right),
\end{equation}
so that the return series \(\{\mathtt{r}_t\}\) is defined for all trading days after the first observation.

To analyze return dynamics in a discrete state space, we discretize \(\{\mathtt{r}_t\}\) into five quintile-based states \(\mathtt{R}_1,\mathtt{R}_2,\mathtt{R}_3,\mathtt{R}_4,\mathtt{R}_5\). For each index separately, we compute the empirical quintile cutoffs \(q_{0.2}, q_{0.4}, q_{0.6}, q_{0.8}\) from the full-sample distribution of \(\{\mathtt{r}_t\}\) over January 2000 to April 2025. Each trading day \(t\) is then assigned to exactly one state according to
\[
\mathtt{R}_1\!:\!\mathtt{r}_t\!\le\!q_{0.2},\quad
\mathtt{R}_2\!:\!q_{0.2}\!<\!\mathtt{r}_t\!\le\!q_{0.4},\quad
\mathtt{R}_3\!:\!q_{0.4}\!<\!\mathtt{r}_t\!\le\!q_{0.6},\quad
\mathtt{R}_4\!:\!q_{0.6}\!<\!\mathtt{r}_t\!\le\!q_{0.8},\quad
\mathtt{R}_5\!:\!\mathtt{r}_t\!>\!q_{0.8}.
\]
$\mathtt{R}_1$ contains the lowest $20\%$ of returns, which correspond to the most negative outcomes in the sample, $\mathtt{R}_2$ contains the next $20\%$, and so on up to $\mathtt{R}_5$, which contains the highest $20\%$ of returns, corresponding to the most positive outcomes. Thus, $\mathtt{R}_1$ and $\mathtt{R}_5$ represent the lower and upper tails of the unconditional return distribution for that index, while $\mathtt{R}_3$ represents the central part. We use fixed full-sample quintile cutoffs for each index so that regime-specific state probabilities and transition behavior are comparable across regimes.

\subsection{Brock - Dechert - Scheinkman test}
Brock–Dechert–Scheinkman (BDS) test is a non-parametric method of testing for nonlinear patterns in time series. This test has its origins in deterministic nonlinear dynamics and chaos theory~\cite{broock1996test,grassberger1983measuring}. 

The null hypothesis is that data in a time series is independently and identically distributed (iid). According to Takens~\cite{takens2006detecting}, the method of delays can be used to embed a scalar time series $\{x_i\}, i=1, 2, 3, ...,N$ into a m-dimensional space as
follows:
\begin{equation}
\Vec{x}{_i} =(x_i, x_{i+t},\dots,x_{i+(m-1)t}), \quad \Vec{x}{_i}\in\mathbb{R}^m
\end{equation}
where $t$ is the index lag.

Correlation integral measures the fractal dimension of deterministic data, i.e., the frequency with which temporal patterns are repeated in the data~\cite{grassberger1983measuring}. The correlation integral at the embedding dimension m is given by

\begin{equation}
C(m,N,r)=
\frac{2}{M(M-1)}\sum_{1\le i<j\le M}
\Theta\bigl(r - \|\Vec{x}{_i} - \Vec{x}{_j}\|\bigr), \quad r>0,
\end{equation}
 where 
 \begin{equation}
     \Theta(a)=
\begin{cases}
0, & a<0 \\
1, & a\ge0.
\end{cases}
 \end{equation}

Here, $N$ is the size of the data sets, $M=N-(m-1)t$ is the number of embedded points in m-dimensional space, and $\|\cdot\|$ denotes the sup-norm. $C(m,N,r)$ measures the fraction of the pairs of points $\Vec{x}{_i}, i=1, 2, 3, ...,M$, whose sup-norm separation is not greater than r. If the limit of $C(m,N,r)$ as $N\rightarrow \infty$ exists for each r, we write of all state vector points that are within r of each other as $C(m,r)=\lim_{N\to\infty}C(m,N,r)$.

If the data is generated by a strictly stationary stochastic process that is absolutely regular, then this limit exists:

\begin{equation}\label{eq:corr_integral_limit}
C(m,r)
=
\iint
\Theta\bigl(r - \|\Vec{x} - \Vec{y}\|\bigr)\,
dF(\Vec{x})\,dF(\Vec{y}), \quad r>0
\end{equation}

When the process is iid, and since $\Theta\bigl(r - \|\Vec{x} - \Vec{y}\|\bigr)\ = \displaystyle \prod_{k=1}^{m} \Theta\bigl(r - |x_k - y_k|\bigr)$, it implies that $C(m,r) = C^m(1,r)$. Also, 
$C(m,r) - C^m(1,r)$ has asymptotic normal distribution, with zero mean and variance as follows:

\begin{equation}
\begin{split}
\frac{\sigma^2(m,M,r)}{4} &= 
m(m-1)\,C^{2(m-1)}\bigl(K-C^2\bigr)
+ K^m - C^{2m} \\[6pt]
&\quad
+2\sum_{i=1}^{m-1}\Bigl[
C^{2i}\bigl(K^{m-i}-C^{2(m-i)}\bigr)
- m\,C^{2(m-i)}\bigl(K-C^2\bigr)\Bigr]
\end{split}
\end{equation}

We can consistently estimate the constants C by C(1, r) and K by 
\begin{equation}
K(m,N,r)
=
\frac{6}{M(M-1)(M-2)}
\sum_{1\le i<j\le M}
\bigl[\Theta\bigl(r-\|\Vec{x}{_i}-\Vec{x}{_j}\|\bigr)\Theta\bigl(r-\|\Vec{x}{_i}-\Vec{x}{_j}\|\bigr)\bigr]
\end{equation}

Under the (null) iid hypothesis, the BDS statistic for $m>1$ is defined as
\begin{equation}
\mathrm{BDS}(m,N,r)
= \frac{\sqrt{M}}{\sigma} \bigl[ 
C(m, r) - C^m (1, r)\bigr]
\end{equation}

It has a limiting standard normal distribution under the null hypothesis of iid as $M\rightarrow\infty$ and obtains the critical values using the standard normal distribution. With this test, we examine the non-linearity feature in each of the return time series. In practice, the indicator of nonlinearity is the BDS test statistic $\mathrm{BDS}(m,N,r)$, equivalently its associated $p$-value, computed for selected embedding dimensions $m>1$ and distance thresholds $r$. Rejection of the iid null hypothesis, that is, statistically significant $\mathrm{BDS}(m,N,r)$ values across a range of $(m,r)$, is taken as evidence of nonlinear dependence in the return series.

\subsection{Hilbert-Huang Transform and Holo-Hilbert Spectral Analysis}

A non-linear time series can have both amplitude and frequency modulations generated by two different mechanisms: linear additive or nonlinear multiplicative processes~\cite{huang2016holo}. Holo-Hilbert Spectral Analysis (HHSA) accommodates all the processes: additive and multiplicative, intra and inter-mode, stationary and nonstationary, linear and nonlinear interactions~\cite{huang2016holo, nguyen2019unraveling}. With the Holo-Hilbert spectrum (HHS), both the carrier frequencies $\omega_c$ and the amplitude modulation frequencies $\omega_{am}$ can be examined simultaneously, together with amplitude modulation energy.

To obtain the Hilbert spectrum from the Hilbert-Huang Transform (HHT) for regime identification and the HHS from HHSA for regime profiling, we proceed as follows. First, the original signal $x(t)$ is decomposed into Intrinsic Mode Functions (IMFs) $c_j(t)$ using Empirical Mode Decomposition (EMD), and is expressed as
\begin{equation}
x(t) = \sum_{j=1}^{n} c_j(t) + q_n
     = \sum_{j=1}^{n} a_j(t)\cos\theta_j(t) + q_n,
\end{equation}
where $\{c_j(t)\}$ are the first-layer IMFs and $q_n$ is the residual. To avoid confusion in terminology, we refer to the instantaneous frequency obtained from the first-layer EMD as the carrier frequency $\omega_c$. Next, the direct quadrature (DQ) method is applied to estimate instantaneous frequencies and amplitudes of the IMFs~\cite{huang2009instantaneous}. This step yields the time--frequency representation of the signal, namely the Hilbert spectrum, and the squared instantaneous amplitudes provide an instantaneous energy measure.

Since the magnitude of instantaneous energy can differ across indices, we normalize the energy series before applying regime thresholds. Let $IA_j(t)$ denote the instantaneous amplitude of the $j$th IMF from HHT. We compute the raw instantaneous energy as
\begin{equation}
E_{\mathrm{raw}}(t)=\sum_{j=1}^{n} IA_j(t)^2,
\end{equation}
and apply max-normalization,
\begin{equation}
E(t)=\frac{E_{\mathrm{raw}}(t)}{\max_{t}\!\left[ E_{\mathrm{raw}}(t)\right]}.
\end{equation}

so that $E(t)\in[0,1]$ for each index. The regime thresholds are then computed using the sample mean $\mu$ and sample standard deviation $\sigma$ of the normalized series $\{E(t)\}$, and regimes are identified using cutoffs at $\mu+\sigma$ and $\mu+6\sigma$.

To obtain the amplitude function of each IMF as defined by Huang et al.~\cite{huang2009instantaneous,huang2013uniqueness,huang2016holo}, we take the absolute value of the IMF, identify the maxima of the absolute-valued IMF, and assemble the envelope by employing a natural spline through these maxima. The second-layer EMD is then obtained by applying masking EMD to the amplitude function $a_j(t)$, giving
\begin{equation}
a_j(t) = \sum_{k=1}^{m} c_{jk}(t) + Q_{jm}
       = \sum_{k=1}^{m} a_{jk}(t)\cos \Theta_{jk}(t) + Q_{jm},
\end{equation}
where $c_{jk}(t)$ are the second-layer IMFs, $a_{jk}(t)$ are the second-layer amplitude functions, $\Theta_{jk}(t)$ are the second-layer phase functions, and $Q_{jm}$ are the trends of each second-layer IMF. The resulting two-layer expansion can be written as
\begin{equation}
x(t) = \sum_{j=1}^{n}\Bigl[\sum_{k=1}^{m} a_{jk}(t)\cos\Theta_{jk}(t) + Q_{jm}\Bigr]\cos\theta_j(t) + q_n.
\end{equation}

The DQ method is again applied to these second-layer IMFs to determine the instantaneous frequency and amplitude of amplitude modulation, denoted by $\omega_{am}$. The instantaneous frequency and amplitude of this two-layer decomposition are projected to $(\omega_{am},\omega_c,t)$ space to obtain the 3-D HHS, which characterizes cross-frequency dynamics varying with time. To aid interpretability, the 3-D HHS is integrated over time to obtain a 2-D HHS, in which the y-axis represents $\omega_c$, the x-axis represents $\omega_{am}$, and the color intensity represents the amplitude modulation energy. This energy metric is computed as the time average of the squared second-layer amplitude functions $a_{jk}(t)$, namely
\begin{equation}
\frac{1}{T}\int_0^T |a_{jk}(t)|^2\,dt,
\end{equation}
where $T$ denotes the total time period.

The 2-D HHS obtained through HHSA provides a comprehensive representation of cross-frequency dynamics~\cite{huang2016holo, nguyen2019unraveling}, simultaneously capturing carrier frequencies ($\omega_c$) and amplitude-modulation frequencies ($\omega_{am}$). In economic terms, $\omega_c$ reflects the dominant time scale of price adjustments: higher $\omega_c$ indicates faster, more frequent price movements consistent with active trading and rapid information incorporation, whereas lower $\omega_c$ indicates slower, more persistent movements. The modulation frequency $\omega_{am}$ describes how rapidly the strength of these price oscillations varies over time, providing a scale-resolved representation of volatility clustering and the temporal instability of risk; higher $\omega_{am}$ corresponds to rapidly changing volatility intensity, while lower $\omega_{am}$ corresponds to more slowly varying volatility conditions. Crucially, the amplitude-modulation energy quantifies the magnitude of volatility intensity at each $(\omega_c,\omega_{am})$ coordinate. This joint $(\omega_c,\omega_{am})$ representation, together with the associated energy, enables the profiling of market regimes through distinct and economically interpretable volatility signatures.

After the Hilbert–Huang-based regime identification and volatility-signature profiling, we turn to modeling how return categories evolve within each regime. Specifically, we use Variable-length Markov chains to capture intra-regime transition dynamics, as described next.

\subsection{Variable-length Markov chain}

Variable-length Markov chains (VLMC) are sparse high-order Markov chains. They model discrete-valued time series in which short memory is sufficient in some situations, while longer memory is needed in others. A collection of past states that determines the next-step transition probabilities is called a context~\cite{buhlmann1999variable,zanin2022variable}.

Let $X_1, X_2, \cdots, X_n, \cdots$ be a sequence of random variables on a finite state space $S$. The sequence is a VLMC if there is a maximal order $\ell_{\max}$ and a function $\ell: S^{\ell_{\max}}\;\longrightarrow\;\{0,1,\dots,\ell_{\max}\}$ such that for all $n>\ell_{\max}$,

\begin{equation}
\begin{aligned}
&P\Bigl(X_n=x_n \,\Bigm|\, X_{n-1}=x_{n-1},X_{n-2}=x_{n-2},\dots,X_1=x_1\Bigr)\\
&\qquad=
P\Bigl(X_n=x_n \,\Bigm|\, X_{n-1}=x_{n-1},\dots, X_{\,n-\ell(x_{n-\ell_{\max}},\dots,x_{n-1})}
= x_{\,n-\ell(x_{n-\ell_{\max}},\dots,x_{n-1})}\bigr).
\end{aligned}
\end{equation}

In other words, the memory length (order) is variable and given by $\ell(x_{n-\ell_{\max}},\dots,x_{n-1})$. The memory-length function generates a context function $c$ that retains the relevant suffix of the past needed to obtain the conditional distribution. Specifically, $c$ is a function from $\displaystyle S^{\ell_{\max}} \longrightarrow \bigcup_{k=0}^{\ell_{\max}} S^k$ given by
\begin{equation}
c\bigl(x_{n-\ell_{\max}},\dots,x_{n-1}\bigr)
=
\bigl(x_{n-\ell},\dots,x_{n-1}\bigr),
\qquad
\ell=\ell(x_{n-\ell_{\max}},\dots,x_{n-1}).
\end{equation}


The image by $c$ of $S^{\ell_{\max}}$ is the set of contexts of the VLMC, which is entirely specified by $\ell$, with one conditional distribution associated with each unique context~\cite{buhlmann1999variable,zanin2022variable}.

\subsubsection*{Toy example: Interpreting a VLMC context tree and the pruning rule}

A VLMC is conveniently represented by a context tree. Each node is labeled by a return state, and each displayed probability vector gives the conditional distribution of the next-day state. The root node marked by $\ast$ reports unconditional probabilities $P(X_{t+1}=\mathtt{R}_i)$, $i=1,\dots,5$. Nodes one level below the root correspond to conditioning on the most recent observation, Day$-1$. Deeper nodes add older lags. For instance, the path $\mathtt{R}_5 \rightarrow \mathtt{R}_1$ represents the two-day context Day$-2=\mathtt{R}_5$ followed by Day$-1=\mathtt{R}_1$. The context length equals the number of states along the path from the root to the node, excluding the root. Hence, $\ell(\mathtt{R}_1)=1$ and $\ell(\mathtt{R}_5\mathtt{R}_1)=2$.

To illustrate, consider the simplified toy tree below. Each node shows the next-day conditional distribution in the order $(\mathtt{R}_1,\mathtt{R}_2,\mathtt{R}_3,\mathtt{R}_4,\mathtt{R}_5)$:
\begin{center}
\begin{forest}
for tree={
    draw,
    rounded corners,
    edge={-latex},
    node options={align=center,font=\scriptsize},
    s sep=6pt,
    l sep=10pt,
    grow=south
},
[
    {$\ast$\\(0.20,\,0.20,\,0.20,\,0.20,\,0.20)}
    [
        {$\mathtt{R}_1$\\(0.30,\,0.15,\,0.20,\,0.10,\,0.25)}
        [
            {$\mathtt{R}_5$\\(0.48,\,0.02,\,0.10,\,0.10,\,0.30)}
        ]
    ]
    [
        {$\mathtt{R}_3$\\(0.18,\,0.22,\,0.30,\,0.18,\,0.12)}
    ]
]
\end{forest}
\end{center}

The node labeled $\mathtt{R}_1$ corresponds to the one-day context $c=\mathtt{R}_1$ and represents
\[
P(X_{t+1}=\mathtt{R}_i \mid X_t=\mathtt{R}_1)=p(\mathtt{R}_i \mid \mathtt{R}_1), \qquad i=1,\dots,5.
\]
The deeper node $\mathtt{R}_5$ as a child of $\mathtt{R}_1$ corresponds to the two-day context $c=\mathtt{R}_5\mathtt{R}_1$ and represents
\[
P(X_{t+1}=\mathtt{R}_i \mid X_t=\mathtt{R}_1,\;X_{t-1}=\mathtt{R}_5)=p(\mathtt{R}_i \mid \mathtt{R}_5\mathtt{R}_1), \qquad i=1,\dots,5.
\]
Thus, the tree encodes variable memory: for some histories the next-day distribution depends only on Day$-1$, whereas for others a longer suffix is retained because it changes transition probabilities in a statistically meaningful way. In this study, VLMC context trees are estimated using the \textit{mixvlmc} package in R~\cite{mixvlmc}.

In the \textit{mixvlmc} estimation procedure, deeper branches are retained only when the conditional distribution at a candidate child node differs sufficiently from that of its parent. This is enforced via a likelihood-ratio pruning rule based on Kullback--Leibler divergence. Let $\widehat{P}_{c}$ denote the estimated next-state distribution at context $c$ and let $\widehat{P}_{\mathrm{suffix}(c)}$ denote the distribution at the parent context given by the one-step shorter suffix. The branching decision is based on
\[
\Lambda(c)
=
2\,n_c\,
D_{\mathrm{KL}}
\!\left(
\widehat{P}_{c}\;\|\;\widehat{P}_{\mathrm{suffix}(c)}
\right),
\]
where $n_c$ is the number of occurrences of context $c$ in the data. Under standard large-sample arguments, $\Lambda(c)$ is compared to a $\chi^2$ cutoff with degrees of freedom $|S|-1$. In our study, $|S|=5$, hence the reference degrees of freedom are $4$. If $\Lambda(c)$ exceeds the cutoff, the branch is retained and the longer context is kept as a distinct leaf with its own conditional distribution. If not, the branch is pruned and the child inherits the parent distribution, which controls overfitting by retaining only statistically meaningful distributional divergence. This criterion governs all branching decisions in the context trees presented in the Results section.

\subsubsection*{VLMC metrics}
\label{subsec:vlmc_metrics}

To compare regime-specific transition behavior across different effective Markov orders, we define metrics derived from the VLMC context-tree probabilities. Order-$1$ metrics summarize one-day persistence and paired reversals, while order-$k\ge2$ metrics summarize how multi-day contexts modify continuation, tail switching, and the likelihood of extreme moves following calm sequences.

\vspace{4pt}

Order-1 metrics: We define two metrics.\\

1. Self-persistence ($\mathtt{M}_i$): Probability of remaining in state $\mathtt{R}_i$ after one step
\begin{equation}
   \mathtt{M}_i = p_i(\mathtt{R}_i), \qquad i=1,\dots,5.
\end{equation}
Higher $\mathtt{M}_i$ indicates short-run inertia in that return category. In particular, elevated $\mathtt{M}_1$ and $\mathtt{M}_5$ reflect clustering of tail outcomes, consistent with heightened short-horizon tail exposure and tighter risk constraints. By contrast, larger $\mathtt{M}_2$--$\mathtt{M}_4$ suggests stable conditions.\\

2. Reversal intensity ($\mathtt{V}_i$): We measure paired reversals using
\begin{equation}
   \mathtt{V}_1 = \frac{1}{2} \left(p_{5}(\mathtt{R}_1) + p_{1}(\mathtt{R}_5)\right), \qquad
   \mathtt{V}_2 = \frac{1}{2} \left(p_{4}(\mathtt{R}_2) + p_{2}(\mathtt{R}_4)\right).
\end{equation}
Here, $\mathtt{V}_1$ captures tail reversals between $\mathtt{R}_1$ and $\mathtt{R}_5$, while $\mathtt{V}_2$ captures moderate reversals between $\mathtt{R}_2$ and $\mathtt{R}_4$. Economically, larger $\mathtt{V}_1$ indicates stronger tail-to-tail flipping, which is a key feature of turbulent conditions and whipsaw-type corrections. A higher $\mathtt{V}_2$ indicates milder back-and-forth movement around typical trading conditions, consistent with progressive normalization.\\

Order $\ge2$ metrics: We define four metrics.\\

1. Continuation ($\mathtt{C}_k$): $\mathtt{C}_k$ is a count-weighted average of ``run continuation'' probabilities over homogeneous $k$-day runs.
\begin{equation}
  \mathtt{C}_k =
  \sum_{i=1}^5
  \left(
  \frac{n_{\mathtt{R_iR_i\ldots R_i}}}{\sum\limits_{c \in \mathcal{C}_k} n_c}
  \cdot p_i^{\mathtt{R_iR_i\ldots R_i}}
  \right)
  \cdot \mathbbmI_{\{\mathtt{R_iR_i\ldots R_i} \in \mathcal{C}_k\}},
\end{equation}

where $\mathcal{C}_k$ denotes all observed contexts of length $k$, $n_c$ is the number of observations for a context $c \in \mathcal{C}_k$, $n_{\mathtt{R_iR_i\ldots R_i}}$ is the count of the homogeneous context $\mathtt{R_iR_i\ldots R_i}$, and $p_i^{\mathtt{R_iR_i\ldots R_i}}$ is the probability of $\mathtt{R}_i$ after that context, and \(\mathbbm{I}_{\{\mathtt{R_iR_i\ldots R_i} \in \mathcal{C}_k\}}\) denotes the indicator function (1 if the context exists, 0 otherwise). Economically, high $\mathtt{C}_k$ indicates multi-day run persistence, which can prolong stress episodes when the run occurs in tail states. In calmer regimes, continuation primarily reflects persistence within middle states and more orderly dynamics.\\

2. Exhaustion ($\mathtt{E}_k$): $\mathtt{E}_k$ measures the tendency to switch between extremes after homogeneous extreme contexts.
\begin{equation}
\mathtt{E}_k = \frac{1}{2} \left( p_5^{\mathtt{R_1R_1\ldots R_1}} + p_1^{\mathtt{R_5R_5\ldots R_5}} \right),
\end{equation}
where $p_5^{\mathtt{R_1R_1\ldots R_1}}$ and $p_1^{\mathtt{R_5R_5\ldots R_5}}$ are the probabilities of switching to $\mathtt{R}_5$ after $k$ consecutive $\mathtt{R}_1$ outcomes and to $\mathtt{R}_1$ after $k$ consecutive $\mathtt{R}_5$ outcomes, respectively. Large $\mathtt{E}_k$ indicates sharp tail-to-tail reversals following $k$-day extreme runs, consistent with abrupt corrections after sustained selling or buying pressure.\\

3. Zigzag Alternation ($\mathtt{Z}_k$): The tendency to alternate between $\mathtt{R}_1$ and $\mathtt{R}_5$ over an alternating context of length $k$, measured as continuation of the alternation pattern:
\begin{equation}
\mathtt{Z}_k = \frac{1}{2} \left(p_5^{\mathtt{R_1R_5R_1\ldots}} + p_1^{\mathtt{R_5R_1R_5\ldots}}\right).
\end{equation}
Here, $\mathtt{R_1R_5R_1\ldots}$ denotes the length-$k$ alternating context that ends in $\mathtt{R}_1$, so $p_5^{\mathtt{R_1R_5R_1\ldots}}$ is the probability that the next state is $\mathtt{R}_5$ (i.e., the alternation continues). Likewise, $\mathtt{R_5R_1R_5\ldots}$ denotes the length-$k$ alternating context that ends in $\mathtt{R}_5$, so $p_1^{\mathtt{R_5R_1R_5\ldots}}$ is the probability that the next state is $\mathtt{R}_1$. Economically, high $\mathtt{Z}_k$ indicates whipsaw markets with rapid sign switching, often associated with low depth, high uncertainty, and frequent liquidity-taking.\\

4. Burst from Calm ($\mathtt{B}_k$): $\mathtt{B}_k$ measures burst of extreme returns $\mathtt{R}_1$ and $\mathtt{R}_5$ after calm contexts built from $\mathtt{R}_2$, $\mathtt{R}_3$ and $\mathtt{R}_4$.
\begin{equation}
\mathtt{B}_k =
\sum_{\substack{c \in \mathcal{C}_k^{\text{calm}}}}
\left(
\frac{n_c}{\sum\limits_{c' \in \mathcal{C}_k} n_{c'}}
\cdot \left( p_1^c + p_5^c \right)
\right),
\end{equation}
where \(\mathcal{C}_k^{\text{calm}} \subset \mathcal{C}_k\) denotes the calm contexts (composed exclusively of returns \(\mathtt{R}_2, \mathtt{R}_3, \mathtt{R}_4\)) of length \(k\), 
\(n_c\) is the number of observations for a specific calm context \(c \in \mathcal{C}_k^{\text{calm}}\), 
\(\sum\limits_{c' \in \mathcal{C}_k} n_{c'}\) is the total number of observations for all contexts of length \(k\) (calm or not), 
\(p_1^c\) is the probability of extreme negative return \(\mathtt{R}_1\) occurring after context \(c\), 
and \(p_5^c\) is the probability of extreme positive return \(\mathtt{R}_5\) occurring after context \(c\). Economically, high $\mathtt{B}_k$ indicates that tail events can emerge directly from apparently stable conditions, consistent with latent fragility, news shocks, or sudden liquidity withdrawal.

\begin{figure}[htbp]
    \centering
    \resizebox{1.0\textwidth}{!}{%
    \providecommand{\NodeBulletList}[1]{%
      \begin{minipage}{\linewidth}
        \begin{list}{\textbullet}{%
          \setlength{\leftmargin}{0.9em}
          \setlength{\itemsep}{0pt}
          \setlength{\topsep}{2pt}
          \setlength{\parsep}{0pt}
          \setlength{\partopsep}{0pt}
        }
          #1
        \end{list}
      \end{minipage}
    }%
    \begin{tikzpicture}[
        node distance=0.8cm and 1.2cm,
        process_data/.style={
            rectangle,
            rounded corners,
            draw=cyan!60!black,
            fill=white,
            thick,
            align=center,
            font=\scriptsize
        },
        process_model/.style={
            rectangle,
            rounded corners,
            draw=orange!60!black,
            fill=white,
            thick,
            align=center,
            font=\scriptsize
        },
        process_analysis/.style={
            rectangle,
            rounded corners,
            draw=green!60!black,
            fill=white,
            thick,
            align=center,
            font=\scriptsize
        },
        arrow/.style={
            -{Latex[length=3mm]},
            thick,
            gray!50!black
        },
        line/.style={
            thick,
            gray!50!black
        },
        phase label/.style={
            font=\bfseries\footnotesize,
            text=black!70
        }
    ]

        \pgfdeclarelayer{background}
        \pgfsetlayers{background,main}

        \node (raw) [process_data, text width=6.0cm] {
            \textbf{Daily Return Data -- 2000 to 2025}\\
            (20 Indices: 10 Developed, 10 Developing)
        };

        \node (bds) [
            process_data,
            below=0.6cm of raw,
            text width=5.0cm,
            inner sep=5pt
        ] {
            \textbf{BDS Test}\\
            (Confirm Non-linearity of Returns)
        };

        \node (hht) [
            process_data,
            below=0.6cm of bds,
            text width=8.0cm,
            inner sep=5pt
        ] {%
            \centering\textbf{Regime Identification with HHT}\\[1pt]
            \begin{minipage}{8.2cm}
            \begin{list}{\textbullet}{%
                \setlength{\leftmargin}{0.9em}
                \setlength{\itemsep}{0pt}
                \setlength{\topsep}{2pt}
                \setlength{\parsep}{0pt}
                \setlength{\partopsep}{0pt}
            }
                \item \raggedright EMD on daily returns $\to$ Intrinsic Mode Functions (IMFs).
                \item \raggedright Hilbert Transform on IMFs $\to$ Inst. Energy (IE).
                \item \raggedright Identify \textbf{Extreme, High, Normal} regimes via IE thresholds $\mu+\sigma$ \& $\mu+6\sigma$.
            \end{list}
            \end{minipage}
        };

        \node (hhsa) [
            process_data,
            below=0.6cm of hht,
            text width=8.3cm,
            inner sep=5pt
        ] {%
            \centering\textbf{Regime Characterization with HHSA}\\[1pt]
            \begin{minipage}{8.9cm}
            \begin{list}{\textbullet}{%
                \setlength{\leftmargin}{0.9em}
                \setlength{\itemsep}{0pt}
                \setlength{\topsep}{2pt}
                \setlength{\parsep}{0pt}
                \setlength{\partopsep}{0pt}
            }
                \item \raggedright $2^{\text{nd}}$-layer EMD on HHT IMFs $\to$ $2^{\text{nd}}$-layer IMFs.
                \item \raggedright Hilbert Transform on $2^{\text{nd}}$-layer IMFs $\to$ AM Energy (AME).
                \item \raggedright Characterize regimes via AME as \textbf{Volatility Intensity}.
            \end{list}
            \end{minipage}
        };


        \node (discretize) [
            process_model,
            below=1.2cm of hhsa,
            text width=6.5cm,
            inner sep=5pt
        ] {%
            \centering\textbf{Return State Discretization}\\[1pt]
            \begin{minipage}{7.4cm}
            \raggedright Discretize daily returns into quintiles -- 5 states.\\
            $\mathtt{R}_1$ (Extreme Loss) $\dots$ $\mathtt{R}_5$ (Extreme Gain)
            \end{minipage}
        };

        \node (vlmc) [
            process_model,
            below=0.6cm of discretize,
            text width=8.9cm,
            inner sep=5pt
        ] {%
            \centering\textbf{Variable-Length Markov Chain (VLMC)}\\[1pt]
            \NodeBulletList{%
                \item \raggedright Transitions between states within each regime via context trees.
                \item \raggedright Compare intra-regime return dynamics for developed \& developing markets.
            }%
        };


        \node (uncond) [
            process_analysis,
            below=1.3cm of vlmc,
            xshift=-3.5cm,
            text width=3.5cm,
            inner sep=5pt
        ] {%
            \centering\textbf{Unconditional Analysis}\\[1pt]
            \NodeBulletList{%
                \item \raggedright State probabilities.
                \item \raggedright Tail ratio.
                \item \raggedright Shannon entropy.
            }%
        };

        \node (cond) [
            process_analysis,
            below=1.3cm of vlmc,
            xshift=3.5cm,
            text width=6.2cm,
            inner sep=5pt
        ] {%
            \centering\textbf{Conditional Dynamics Metrics}\\[1pt]
            \NodeBulletList{%
                \item \raggedright \textbf{Order 1}: Self-persistence, reversal intensity.
                \item \raggedright \textbf{Order 2 \& 3}: Continuation, exhaustion, zigzag alternation, burst from calm.
            }%
        };


        \draw[arrow] (raw) -- (bds);
        \draw[arrow] (bds) -- (hht);
        \draw[arrow] (hht) -- (hhsa);
        \draw[arrow] (hhsa) -- (discretize);
        \draw[arrow] (discretize) -- (vlmc);

        \coordinate (fork) at ($(vlmc.south) + (0,-0.65cm)$);
        \draw[line] (vlmc.south) -- (fork);
        \draw[arrow] (fork) -| (uncond.north);
        \draw[arrow] (fork) -| (cond.north);

        \begin{pgfonlayer}{background}
            \node[fit=(raw)(bds)(hht)(hhsa),
                  fill=cyan!5,
                  rounded corners,
                  draw=cyan!30,
                  dashed,
                  inner sep=12pt] (p1) {};
            \node[phase label, anchor=north] at (p1.north) {Regime Discovery \& Characterization};

            \node[fit=(discretize)(vlmc),
                  fill=orange!5,
                  rounded corners,
                  draw=orange!30,
                  dashed,
                  inner sep=10pt] (p2) {};
            \node[phase label, anchor=north, xshift=-2.0cm, yshift=0.1cm] at (p2.north) {Dynamics Modeling};

            \node[fit=(uncond)(cond),
                  fill=green!5,
                  rounded corners,
                  draw=green!30,
                  dashed,
                  inner sep=10pt] (p3) {};
            \node[phase label, anchor=north, xshift=-0.4cm, yshift=0.1cm] at (p3.north) {Comparative Analysis};
        \end{pgfonlayer}

    \end{tikzpicture}
    } 
   \caption{Methodological flowchart illustrating the pipeline of the study: Daily returns data, checking non-linearity via BDS test, identifying and profiling market regimes via Empirical mode decomposition (EMD)-based Hilbert--Huang Transform (HHT) and Holo-Hilbert Spectral Analysis (HHSA), followed by intra-regime return dynamics modeling using Variable-Length Markov Chains (VLMC) and analysis with metrics.}
    \label{fig:methodology_flowchart}
\end{figure}
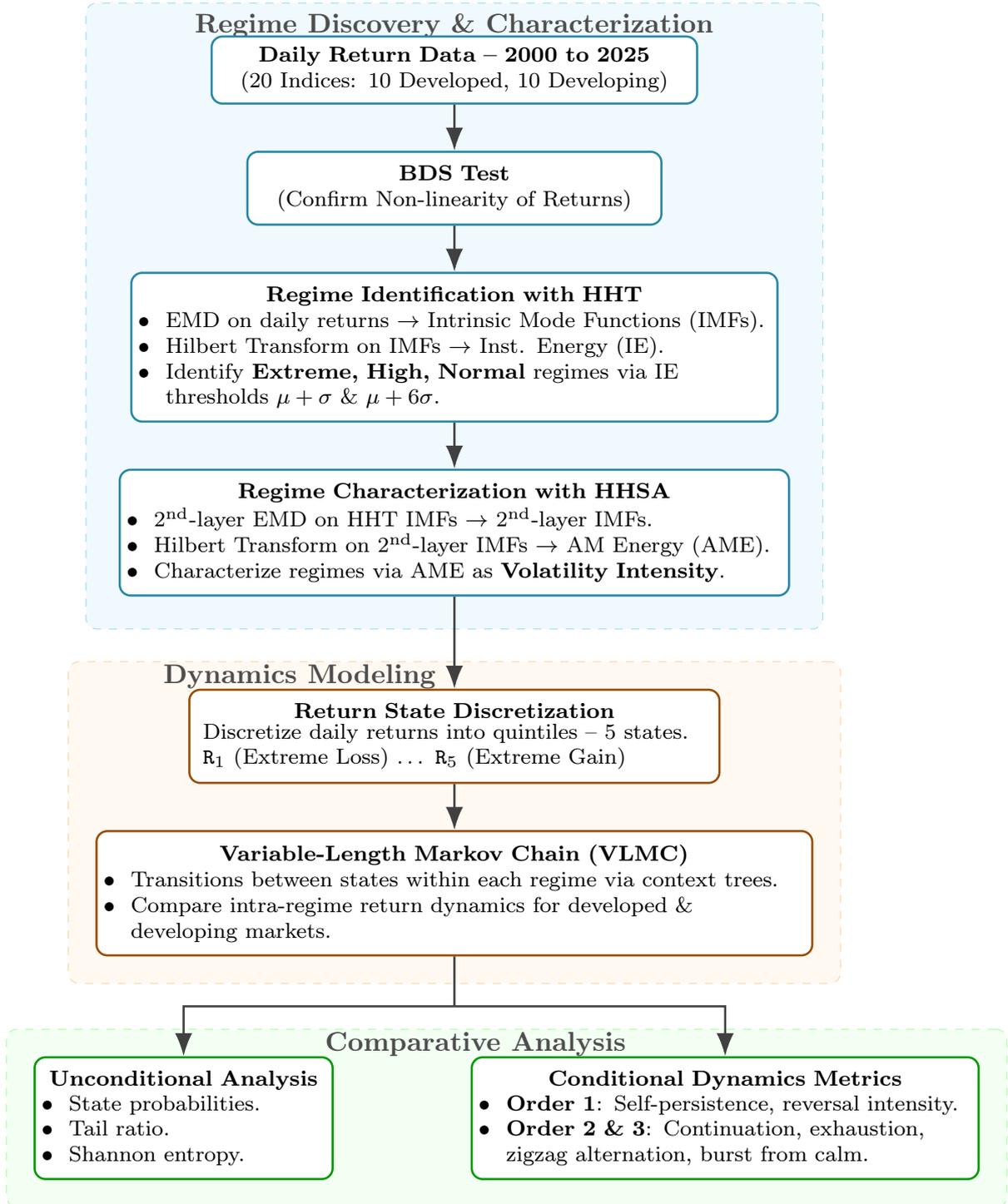

Figure~\ref{fig:methodology_flowchart} summarizes the complete methodological framework, detailing the progression from regime identification and characterization to the modeling of intra-regime return dynamics.

\section{Results}
\label{sec:R}

In this section, we present the results of our study. We begin in Subsection~\ref{R:bds} by applying the Brock--Dechert--Scheinkman test to confirm the nonlinear nature of the return time series. We then identify three market regimes for all indices in both developed and developing markets within an Empirical Mode Decomposition-based Hilbert--Huang Transform framework. Subsection~\ref{R:hhsa} uses instantaneous energy from the Hilbert spectrum to separate Normal, High, and Extreme regimes. For each one-year regime segment, it reports the corresponding Holo--Hilbert spectrum (HHS). The HHS jointly resolves carrier frequencies associated with price movements and amplitude-modulation frequencies capturing volatility fluctuations. This representation highlights regime-wise modulation-energy differences. Finally, Subsection~\ref{R:vlmc} investigates regime-specific transition dynamics among discretized daily return states using a variable-length Markov chain (VLMC) approach, enabling a comparative assessment of return-state dependence across regimes and between developed and developing markets.

\subsection{BDS test}
\label{R:bds}

The Brock--Dechert--Scheinkman (BDS) test is carried out to examine departures from independent and identically distributed behavior in the daily return series. Table~\ref{tab:bds} in the Appendix reports the BDS test statistics and corresponding $p$-values for embedding dimensions $m=2$ and $m=3$. We focus on these low dimensions because, in finite samples, the BDS statistic becomes less stable and less powerful as $m$ increases due to the rapid sparsity of close pairs in higher-dimensional embeddings~\cite{broock1996test,brock1991nonlinear}. Using $m=2$ and $m=3$ therefore provides a reliable and widely used diagnostic for nonlinear dependence while avoiding over-embedding. For the neighborhood size, we set $\varepsilon = 0.5\sigma$, where $\sigma$ is the sample standard deviation of returns. This follows the common practice of scaling $\varepsilon$ by $\sigma$ so that the neighborhood is comparable across indices and remains sufficiently local to retain good power without making the correlation integral too sparse~\cite{broock1996test,hsieh1991chaos}.

For both developed and developing markets, every index is significant at the $5\%$ level at one or both embedding dimensions, with the exception of BVSP whose $p$-values of $0.42$ ($m=2$) and $0.08$ ($m=3$) provide no evidence against iid behavior at $\varepsilon = 0.5\sigma$. Overall, daily returns for almost all indices display clear nonlinear dependence. Therefore, the subsequent analyses employ methods that accommodate nonlinear features in return dynamics.

\subsection{Regime identification and profiling}
\label{R:hhsa}

Accommodating the non-linear feature of the return time-series of the indices, Empirical mode decomposition-based Hilbert--Huang Transform (HHT) and Holo-Hilbert Spectral Analysis (HHSA) are used in this study for regime identification and profiling the identified regimes, respectively.

\begin{figure}[]
  \centering
  \includegraphics[width=0.9\linewidth]{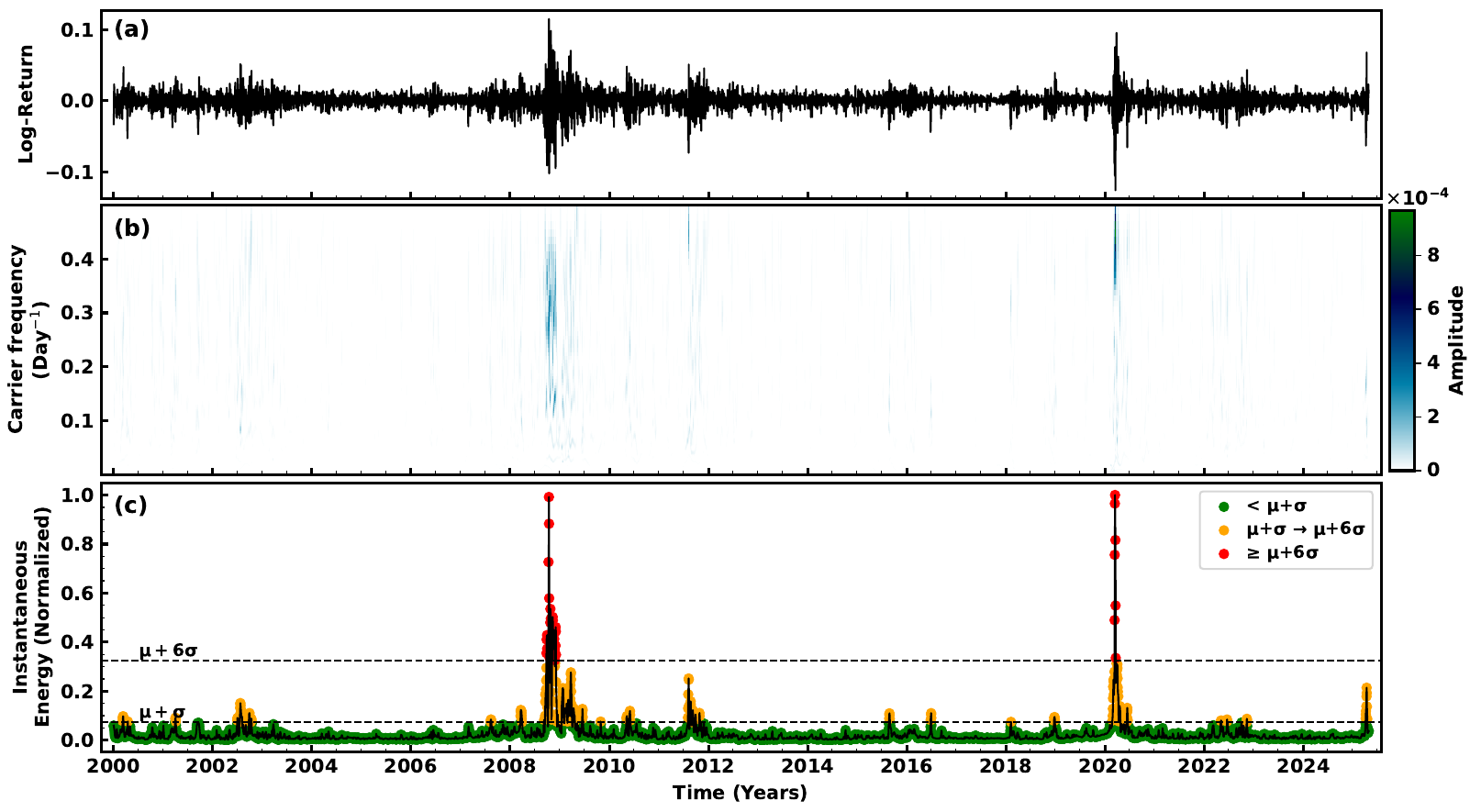}
  \caption{Regime classification for the NYSE Composite index NYA. Panel (a) shows the daily log-returns of the closing price. Panel (b) shows the 2D Hilbert spectrum from the Hilbert--Huang Transform, with carrier frequency on the vertical axis and time on the horizontal axis, and color indicating amplitude. Panel (c) shows the normalized instantaneous energy $E(t)$ computed from the instantaneous amplitudes associated with the Hilbert spectrum. Points are color-coded using energy thresholds, with green denoting Normal for $E(t)\le \mu+\sigma$, orange denoting High for $\mu+\sigma < E(t)\le \mu+6\sigma$, and red denoting Extreme for $E(t)>\mu+6\sigma$. Dashed horizontal lines mark $\mu+\sigma$ and $\mu+6\sigma$, where $\mu$ and $\sigma$ are the sample mean and standard deviation of the normalized energy series.}
  \label{fig:HHT_IE}
\end{figure}

Figures~\ref{fig:HHT_IE}(a) -- (c) show the daily log returns of the NYSE Composite Index (NYA), its Hilbert spectrum from HHT, and the corresponding instantaneous energy plot. Based on the statistical thresholds~\cite{rai2023detection} of the instantaneous energy distribution, we classify three distinct market regimes: (a) Extreme [$E(t)>\mu + 6 \sigma$] marked by red points, (b) High [$ \mu + \sigma<E(t)\leq\mu + 6 \sigma$] marked by orange points, and (c) Normal [$E(t)\leq\mu + \sigma$] marked by green points. Here, $E(t)$ denotes the normalized instantaneous energy at time $t$, and $\mu$ and $\sigma$ represent its sample mean and standard deviation, respectively. 

For the Extreme regime, the extreme movements are concentrated within a few months of 2008 and 2020. However, we select one-year windows representative of each regime in order to capture the broader market dynamics including anticipatory moves preceding the extreme movements and aftershock effects following them~\cite{lillo2003power,scheffer2009early,rai2022statistical}. To ensure consistent analysis and comparability across all regimes, we similarly consider both the High and Normal regimes for one-year periods. Table~\ref{tab:regime_years_developed} presents the identified one-year periods for developed market indices. For this group, the representative years consistently associated with each regime are 2008 and 2020 for Extreme, 2015 and 2022 for High, and 2005 and 2017 for Normal. Similarly, Table~\ref{tab:regime_years_developing} displays the corresponding periods for developing markets. The common regime years are 2008 and 2020 for Extreme, 2002 and 2004 for High, and 2017 and 2023 for Normal.

\begin{figure}[H]
  \centering
  \begin{subfigure}[b]{0.32\linewidth}
    \centering
    \includegraphics[width=\linewidth]{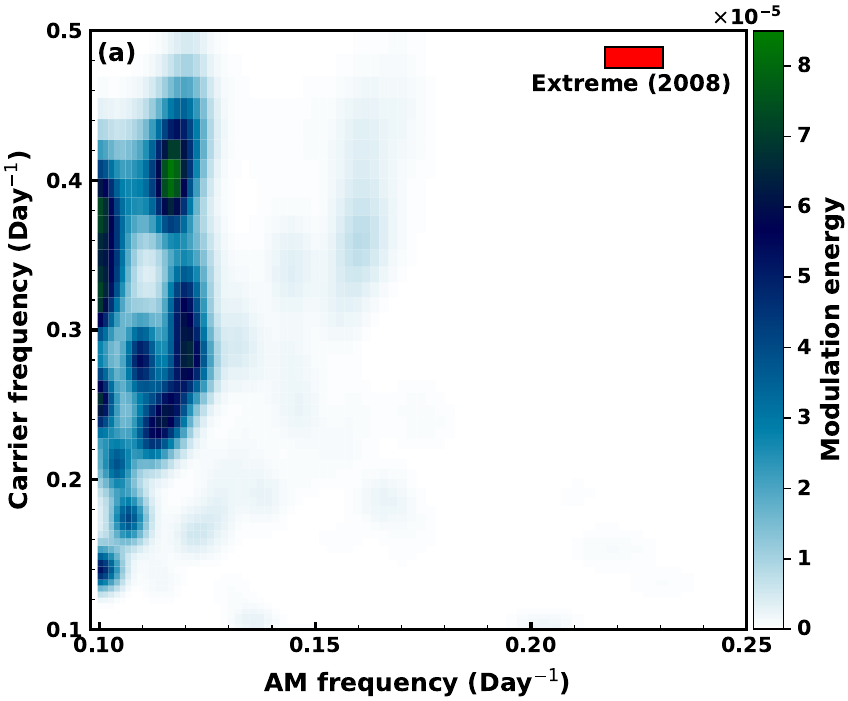}
    \label{fig:2008}
  \end{subfigure}\hfill
  \begin{subfigure}[b]{0.32\linewidth}
    \centering
    \includegraphics[width=\linewidth]{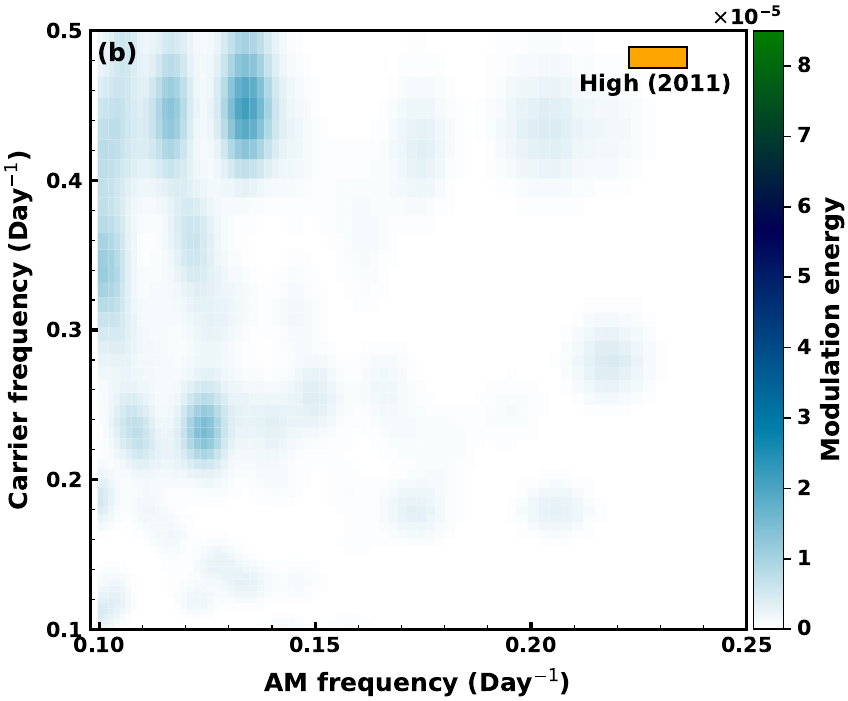}
    \label{fig:2011}
  \end{subfigure}\hfill
  \begin{subfigure}[b]{0.32\linewidth}
    \centering
    \includegraphics[width=\linewidth]{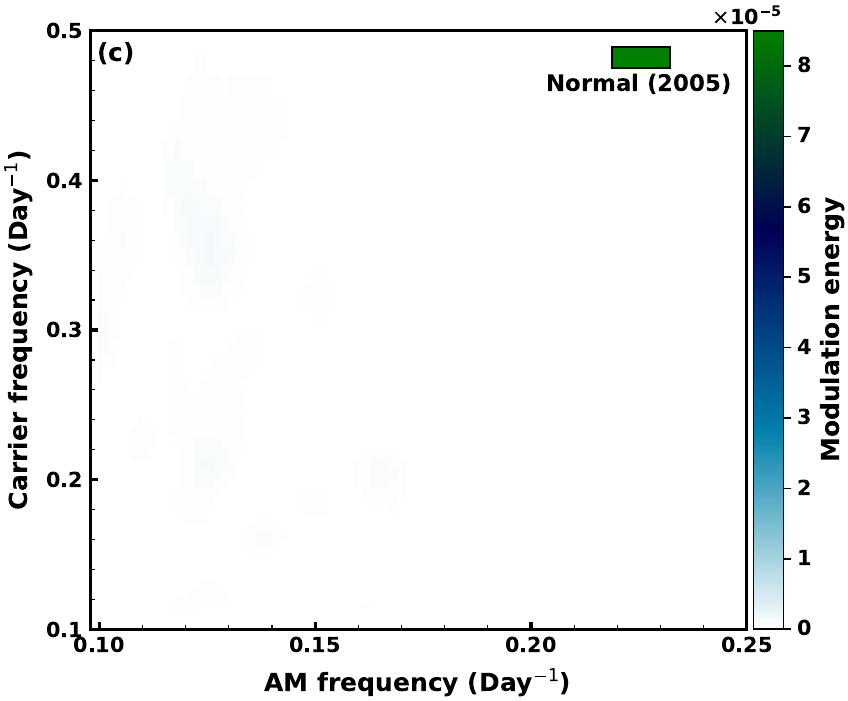}
   \label{fig:2005}
  \end{subfigure}
  
  \caption{Holo--Hilbert spectra (HHS) for the NYSE Composite index (NYA) over one-year windows selected to represent the three regimes identified from the instantaneous energy series in Fig.~\ref{fig:HHT_IE}: (a) Extreme regime year 2008, corresponding to the red-coded energy points, (b) High regime year 2011, corresponding to the orange-coded energy points, and (c) Normal regime year 2005, corresponding to the green-coded energy points. In each panel, the vertical axis is the carrier frequency $\omega_c$ and the horizontal axis is the amplitude-modulation frequency $\omega_{am}$, while the color scale indicates amplitude-modulation energy (volatility intensity).}
  \label{fig:HHS_panels}
\end{figure}

Figures~\ref{fig:HHS_panels}(a) -- (c) compare the Holo-Hilbert spectrum (HHS)-based volatility signatures across the three market regimes, thereby profiling each regime's cross-frequency volatility structure, with NYA as an illustrative example. The corresponding HHS panels for BVSP are provided in the Appendix~\ref{fig:HHS_panels_BVSP} to illustrate the same Extreme--High--Normal contrast for a representative developing index. In each HHS, the y-axis represents the carrier frequencies ($\omega_c$), while the x-axis represents the amplitude modulation frequencies ($\omega_{am}$). The color intensity depicts the amplitude modulation energy, indicating the magnitude of volatility intensity at each $(\omega_c,\omega_{am})$ coordinate. From these HHS, we observed that the volatility intensity sharply decreases from Extreme to High to Normal regimes. Due to this sharp decrease, the energies and corresponding ($\omega_c$, $\omega_{am}$) pairs are not clearly visible in the High and Normal regimes. To enable meaningful cross-market comparison, we profile each regime numerically using peak amplitude modulation energy (PAME) values and the $95^{\text{th}}$ percentile carrier and amplitude-modulation frequencies ($\omega_c$, $\omega_{am}$), presented numerically in Tables~\ref{tab:hhs_developed} and \ref{tab:hhs_developing} for developed and developing markets respectively.


\begin{table}[H]
\centering
\footnotesize
\caption{Peakamplitude modulation energy (PAME, $\times10^{-5}$), $95^{\text{th}}$-percentile carrier frequency ($\omega_{c}$) and $95^{\text{th}}$-percentile amplitude-modulated frequency ($\omega_{am}$)—for developed-market indices under Extreme, High and Normal regimes.}
\setlength{\tabcolsep}{7pt}
\renewcommand{\arraystretch}{1.3}
\begin{tabular}{|l|r|r|r|r|r|r|r|r|r|}
\hline
\multirow{2}{*}{Index} &
\multicolumn{3}{c|}{Extreme} &
\multicolumn{3}{c|}{High} &
\multicolumn{3}{c|}{Normal} \\
\cline{2-10}
 & \multicolumn{1}{c|}{PAME} & \multicolumn{1}{c|}{$\omega_{c}$} & \multicolumn{1}{c|}{$\omega_{am}$} 
 & \multicolumn{1}{c|}{PAME} & \multicolumn{1}{c|}{$\omega_{c}$} & \multicolumn{1}{c|}{$\omega_{am}$} 
 & \multicolumn{1}{c|}{PAME} & \multicolumn{1}{c|}{$\omega_{c}$} & \multicolumn{1}{c|}{$\omega_{am}$} \\
\hline
AXJO        & 2.343 & 0.221 & 0.108 & 0.746 & 0.202 & 0.104 & 0.189 & 0.243 & 0.106 \\
BFX         & 3.838 & 0.235 & 0.077 & 1.187 & 0.209 & 0.094 & 0.246 & 0.254 & 0.082 \\
FCHI        & 4.171 & 0.222 & 0.103 & 1.155 & 0.220 & 0.113 & 0.225 & 0.231 & 0.109 \\
FTSE        & 3.274 & 0.194 & 0.120 & 0.607 & 0.251 & 0.125 & 0.178 & 0.224 & 0.120 \\
GDAXI       & 4.988 & 0.230 & 0.125 & 1.280 & 0.221 & 0.122 & 0.309 & 0.239 & 0.116 \\
IBEX        & 5.504 & 0.228 & 0.105 & 1.230 & 0.202 & 0.101 & 0.401 & 0.234 & 0.108 \\
KS11        & 6.264 & 0.211 & 0.065 & 0.937 & 0.220 & 0.074 & 0.675 & 0.226 & 0.080 \\
N225        & 7.599 & 0.249 & 0.115 & 1.448 & 0.256 & 0.127 & 0.715 & 0.208 & 0.119 \\
NYA         & 7.218 & 0.308 & 0.115 & 1.003 & 0.194 & 0.118 & 0.130 & 0.246 & 0.127 \\
SSMI        & 3.662 & 0.167 & 0.127 & 1.060 & 0.251 & 0.132 & 0.166 & 0.230 & 0.128 \\
\hline
\textbf{Average} & \textbf{4.886} & \textbf{0.227} & \textbf{0.106} & \textbf{1.065} & \textbf{0.223} & \textbf{0.111} & \textbf{0.324} & \textbf{0.234} & \textbf{0.110} \\
\hline
\end{tabular}
\label{tab:hhs_developed}
\end{table}

Table~\ref{tab:hhs_developed} highlights a sharp reduction in modulation energy for developed markets as they shift from Extreme to Normal regimes. The average $\mathrm{PAME}$ plummets from $4.89\times10^{-5}$ in the Extreme regime to just $0.32\times10^{-5}$ in the Normal regime, accompanied by only modest shifts in $\omega_c$ and $\omega_{am}$. In contrast, Table~\ref{tab:hhs_developing} indicates a smoother decline for developing markets. Here, the $\mathrm{PAME}$ decreases from $4.08\times10^{-5}$ in the Extreme to $0.61\times10^{-5}$ in the Normal. Notably, while the energy drops in both cases, developing markets maintain a significantly higher baseline energy in the Normal regime compared to developed markets.

\begin{table}[H]
\centering
\footnotesize
\caption{Peak amplitude modulation energy (PAME, $\times10^{-5}$), $95^{\text{th}}$-percentile carrier frequency ($\omega_{c}$) and $95^{\text{th}}$-percentile amplitude-modulated frequency ($\omega_{am}$)—for developing-market indices under Extreme, High and Normal regimes.}
\setlength{\tabcolsep}{7pt}
\renewcommand{\arraystretch}{1.3}
\begin{tabular}{|l|r|r|r|r|r|r|r|r|r|}
\hline
\multirow{2}{*}{Index} &
\multicolumn{3}{c|}{Extreme} &
\multicolumn{3}{c|}{High} &
\multicolumn{3}{c|}{Normal} \\
\cline{2-10}
 & \multicolumn{1}{c|}{PAME} & \multicolumn{1}{c|}{$\omega_{c}$} & \multicolumn{1}{c|}{$\omega_{am}$} 
 & \multicolumn{1}{c|}{PAME} & \multicolumn{1}{c|}{$\omega_{c}$} & \multicolumn{1}{c|}{$\omega_{am}$} 
 & \multicolumn{1}{c|}{PAME} & \multicolumn{1}{c|}{$\omega_{c}$} & \multicolumn{1}{c|}{$\omega_{am}$} \\
\hline
BVSP         & 9.393 & 0.234 & 0.110 & 2.150 & 0.233 & 0.104 & 1.155 & 0.255 & 0.103 \\
JKSE         & 3.115 & 0.259 & 0.102 & 1.010 & 0.256 & 0.115 & 0.312 & 0.231 & 0.111 \\
MERV         & 4.873 & 0.208 & 0.148 & 1.790 & 0.210 & 0.147 & 1.303 & 0.199 & 0.139 \\
MXX          & 6.011 & 0.208 & 0.064 & 1.800 & 0.208 & 0.054 & 0.498 & 0.223 & 0.068 \\
SET.BK       & 7.492 & 0.185 & 0.081 & 0.991 & 0.238 & 0.096 & 0.232 & 0.203 & 0.081 \\
STI          & 2.873 & 0.211 & 0.065 & 0.901 & 0.198 & 0.077 & 0.190 & 0.211 & 0.059 \\
TASI.SR      & 0.009 & 0.109 & 0.273 & 0.001 & 0.183 & 0.263 & 0.001 & 0.025 & 0.266 \\
TWII         & 1.835 & 0.233 & 0.088 & 1.710 & 0.216 & 0.081 & 0.272 & 0.244 & 0.080 \\
000001.SS    & 1.061 & 0.220 & 0.155 & 0.422 & 0.206 & 0.165 & 0.085 & 0.252 & 0.167 \\
0388.HK      & 4.157 & 0.232 & 0.123 & 1.990 & 0.249 & 0.122 & 2.029 & 0.227 & 0.122 \\
\hline
\textbf{Average} & \textbf{4.082} & \textbf{0.210} & \textbf{0.121} & \textbf{1.276} & \textbf{0.220} & \textbf{0.122} & \textbf{0.608} & \textbf{0.207} & \textbf{0.120} \\
\hline
\end{tabular}
\label{tab:hhs_developing}
\end{table}

Interpreting these numerical profiles, we observe that while both market types show reduced volatility intensity as measured by amplitude modulation energy moving from Extreme to Normal regimes, the contrast is significantly more pronounced in developed markets. The spectral parameters -- ($\omega_c$), and ($\omega_{am}$) further highlight fundamental structural differences. In developed markets, price movements are fastest in the Normal regime while volatility fluctuations are slowest, a pattern consistent with deeper liquidity and more efficient price discovery in stable periods. In contrast, developing markets exhibit their fastest price movements in the High regime with persistently faster volatility fluctuations across all regimes, suggesting a greater sensitivity to external shocks that transmit volatility more readily. Even in Normal regimes, developing markets maintain substantially higher baseline $\mathrm{PAME}$, i.e. more frequent volatility fluctuations than developed markets. These variations -- price dynamics ($\omega_c$) and volatility behavior ($\omega_{am}$), along with volatility intensity ($\mathrm{PAME}$) -- point to divergent regime-dependent dynamics that are strongly conditioned by market maturity, whether a market is developed or developing.


\subsection{Regime-dependent return dynamics}
\label{R:vlmc}

Following HHT-based regime identification, HHSA-based profiling reveals distinct volatility signatures across Extreme, High, and Normal regimes, establishing clear regime dependence in the underlying return environment. We now examine how this regime dependence is reflected in the day-to-day evolution of returns by studying intra-regime transition dynamics. To analyze these dynamics, we first categorize daily index returns into quintiles $\mathtt{R}_1, \mathtt{R}_2, \mathtt{R}_3, \mathtt{R}_4 \hspace{2pt} \text{and}\hspace{2pt} \mathtt{R}_5$ based on their magnitude -- where $\mathtt{R}_1$ represents the lowest 20\% of returns and $\mathtt{R}_5$ the highest. With these quintiles as the discrete states, we model regime-specific return transitions using the Variable-length Markov Chain (VLMC) framework and the associated metrics defined in Section~\ref{subsec:vlmc_metrics}.

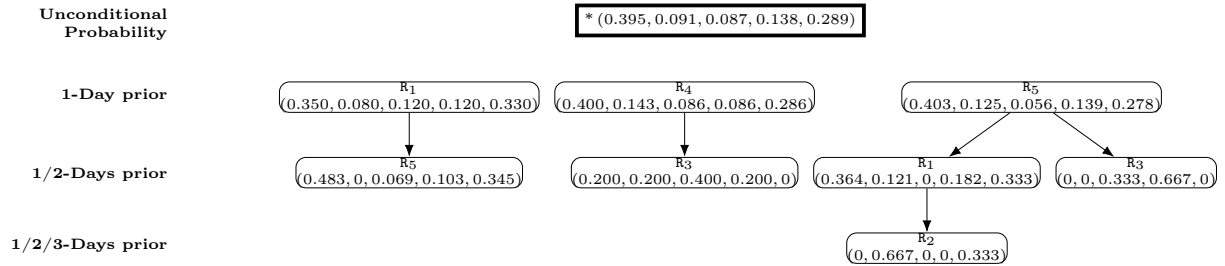
\begin{figure}[H]
\centering
\resizebox{\textwidth}{!}{%
\begin{tikzpicture}[
  daylabel/.style={font=\tiny, anchor=east, align=right},
  transform shape
]
\node (tree) [inner sep=0pt] {
\begin{forest}
  for tree={
    draw,
    rounded corners,
    node options={align=center,font=\tiny},
    edge={-{Latex}}, 
    s sep=0.5em,
    l sep=1.5em,
    anchor=north,
    grow=south,
    where level=1{no edge}{}
  },
  for root={
    rectangle,
    sharp corners,
    line width=1.5pt,
    inner sep=3pt
  }
  [
    {*\,(0.395,\,0.091,\,0.087,\,0.138,\,0.289)}
      [
        {$\mathtt{R}_1$\\(0.350,\,0.080,\,0.120,\,0.120,\,0.330)}
          [{$\mathtt{R}_5$\\(0.483,\,0,\,0.069,\,0.103,\,0.345)}]
      ]
      [
        {$\mathtt{R}_4$\\(0.400,\,0.143,\,0.086,\,0.086,\,0.286)}
          [{$\mathtt{R}_3$\\(0.200,\,0.200,\,0.400,\,0.200,\,0)}]
      ]
      [
        {$\mathtt{R}_5$\\(0.403,\,0.125,\,0.056,\,0.139,\,0.278)}
          [
            {$\mathtt{R}_1$\\(0.364,\,0.121,\,0,\,0.182,\,0.333)}
              [{$\mathtt{R}_2$\\(0,\,0.667,\,0,\,0,\,0.333)}]
          ]
          [{$\mathtt{R}_3$\\(0,\,0,\,0.333,\,0.667,\,0)}]
      ]
  ]
\end{forest}
};

\path (tree.north west) ++(-1.4cm,-0.3cm) node[daylabel] {\textbf{Unconditional}\\\textbf{Probability}};
\path (tree.north west) ++(-1.4cm,-1.3cm) node[daylabel] {\textbf{1-Day prior}};
\path (tree.north west) ++(-1.4cm,-2.4cm) node[daylabel] {\textbf{1/2-Days prior}};
\path (tree.north west) ++(-1.4cm,-3.4cm) node[daylabel] {\textbf{1/2/3-Days prior}};
\end{tikzpicture}%
}

\caption{Extreme (2008) regime context tree for NYSE Composite (NYA) index. The root node (*) with a rectangular box in bold border shows unconditional probabilities of $\mathtt{R}_1, \mathtt{R}_2, \mathtt{R}_3, \mathtt{R}_4 \hspace{2pt} \text{and}\hspace{2pt} \mathtt{R}_5$. First-level nodes represent conditioning on the most recent day (Day$-1$). Deeper nodes represent longer context sequences by adding older lags (Day$-2$, Day$-3$, etc.); e.g., the child node $\mathtt{R}_5$ under $\mathtt{R}_1$ corresponds to the two-day context [Day$-2=\mathtt{R}_5$] and [Day$-1=\mathtt{R}_1$].}
    \label{fig:vlmc-case1}
\end{figure}

Figure~\ref{fig:vlmc-case1} displays the context tree for the NYSE Composite (NYA) index during the Extreme regime (2008). The topmost node marked with \(\ast\) and enclosed in a rectangular box represents the unconditional probabilities of return states, explicitly labeled on the left side. Below this root level, subsequent tiers capture conditional probabilities at increasing orders $k$. Throughout, contexts are written from older to newer states. At $k=1$, nodes summarize one-day conditioning on the most recent state. The \(\mathtt{R}_1\) node represents the transition probabilities conditioned solely on the most recent state being \(\mathtt{R}_1\). This gives the conditional distribution:
\begin{align*}
P(\text{next state} \mid \mathtt{R}_1) = 
\begin{cases} 
\mathtt{R}_1: & 35.00\% \\
\mathtt{R}_2: & 8.00\% \\
\mathtt{R}_3: & 12.00\% \\
\mathtt{R}_4: & 12.00\% \\
\mathtt{R}_5: & 33.00\% 
\end{cases}
\end{align*}

This indicates that after an \(\mathtt{R}_1\) day, returns most commonly persist in \(\mathtt{R}_1\) (35\%) or jump to \(\mathtt{R}_5\) (33\%). At $k=2$, nodes represent two-day contexts. The \(\mathtt{R}_5\) child node under \(\mathtt{R}_1\) corresponds to the two-day context \(\mathtt{R}_5\mathtt{R}_1\). This yields a fundamentally different transition distribution compared to the one-day context \(\mathtt{R}_1\):
\begin{align*}
P(\text{next}\mid \mathtt{R}_5\mathtt{R}_1) = 
\begin{cases} 
\mathtt{R}_1: & 48.28\% \\
\mathtt{R}_2: & 0.00\% \\
\mathtt{R}_3: & 6.90\% \\
\mathtt{R}_4: & 10.34\% \\
\mathtt{R}_5: & 34.48\% 
\end{cases}
\end{align*}

Notably, the probability of reverting back to \(\mathtt{R}_1\) increases from 35.0\% to 48.3\%, and transitions to \(\mathtt{R}_2\) become impossible, dropping from 8.0\% to 0\%. Deeper branches in Fig.~\ref{fig:vlmc-case1} correspond to additional two-day and three-day contexts retained by the estimation procedure. In the context tree estimation, we set the likelihood ratio cutoff to 3.372 (corresponding to the 0.15 quantile of the $\chi^2$ distribution) to retain only context branches exhibiting statistically significant distributional divergence. This ensures that we capture only meaningful dynamics while preventing overfitting by pruning insignificant branches. For example, the distinct context \(\mathtt{R}_5\mathtt{R}_1\) exists exclusively because its log-likelihood ratio statistic, \(2\,n_{\text{seq}}\,D_{\mathrm{KL}}\!\bigl(\widehat{P}_{\mathtt{R}_5\mathtt{R}_1}\;\|\;\widehat{P}_{\mathtt{R}_1}\bigr)\), exceeds \(\epsilon_{\chi^{2}}=3.372\), confirming its power to provide statistically significant new information about next-day distributions. On the other hand, contexts sharing the same suffix \(\mathtt{R}_1\) but not exceeding this threshold remain unbranched and inherit their transition distributions directly from the \(\mathtt{R}_1\) node. This likelihood ratio criterion uniformly governs all branching decisions: every node in the tree represents a sequence where transition probabilities diverge significantly from its parent context.

To generalize context patterns, we aggregate all contexts across stock indices, retaining only those with frequency $>2$ and computing their averaged conditional probabilities. Table~\ref{tab:developed-extreme} shows these aggregated contexts during the Extreme regime, for developed stock market indices.

\begin{table}[H]
\centering
\footnotesize
\caption{Contexts with Count $>2$ for developed stock market indices during Extreme regimes.}
\label{tab:developed-extreme}
\begin{tabular}{|l|c|*{5}{>{\centering\arraybackslash}p{1.2cm}|}}
\hline
\multirow{2}{*}{Context} & \multirow{2}{*}{Count} & \multicolumn{5}{c|}{Probability to after (State)} \\
\cline{3-7}
& & P$(\mathtt{R}_1)$ & P$(\mathtt{R}_2)$ & P$(\mathtt{R}_3)$ & P$(\mathtt{R}_4)$ & P$(\mathtt{R}_5)$ \\
\hline
$\mathtt{R}_1$ & 18 & 0.303 & 0.117 & 0.108 & 0.135 & 0.338 \\
$\mathtt{R}_1\mathtt{R}_1$ & 6 & 0.307 & 0.115 & 0.106 & 0.065 & 0.407 \\
$\mathtt{R}_1\mathtt{R}_1\mathtt{R}_1$ & 4 & 0.128 & 0.257 & 0.130 & 0.130 & 0.355 \\
$\mathtt{R}_1\mathtt{R}_3$ & 3 & 0.185 & 0.201 & 0.245 & 0.160 & 0.209 \\
$\mathtt{R}_1\mathtt{R}_4$ & 3 & 0.161 & 0.138 & 0.199 & 0.088 & 0.414 \\
$\mathtt{R}_1\mathtt{R}_5$ & 9 & 0.368 & 0.073 & 0.061 & 0.108 & 0.391 \\
$\mathtt{R}_1\mathtt{R}_5\mathtt{R}_4$ & 3 & 0.063 & 0.167 & 0.188 & 0.417 & 0.167 \\
\hline
$\mathtt{R}_2$ & 12 & 0.284 & 0.174 & 0.135 & 0.165 & 0.242 \\
$\mathtt{R}_2\mathtt{R}_1$ & 4 & 0.588 & 0.067 & 0.092 & 0.148 & 0.104 \\
$\mathtt{R}_2\mathtt{R}_2$ & 4 & 0.520 & 0.088 & 0.073 & 0.257 & 0.061 \\
$\mathtt{R}_2\mathtt{R}_3$ & 4 & 0.177 & 0.377 & 0.070 & 0.176 & 0.200 \\
$\mathtt{R}_2\mathtt{R}_4$ & 4 & 0.084 & 0.117 & 0.285 & 0.415 & 0.100 \\
\hline
$\mathtt{R}_3$ & 12 & 0.285 & 0.186 & 0.147 & 0.165 & 0.216 \\
$\mathtt{R}_3\mathtt{R}_1$ & 4 & 0.451 & 0.057 & 0.099 & 0.179 & 0.214 \\
$\mathtt{R}_3\mathtt{R}_2$ & 4 & 0.175 & 0.163 & 0.192 & 0.229 & 0.242 \\
$\mathtt{R}_3\mathtt{R}_3$ & 6 & 0.150 & 0.114 & 0.313 & 0.087 & 0.336 \\
\hline
$\mathtt{R}_4$ & 13 & 0.321 & 0.232 & 0.126 & 0.115 & 0.206 \\
$\mathtt{R}_4\mathtt{R}_1$ & 4 & 0.510 & 0.178 & 0.026 & 0.144 & 0.143 \\
\hline
$\mathtt{R}_5$ & 13 & 0.319 & 0.172 & 0.111 & 0.136 & 0.262 \\
$\mathtt{R}_5\mathtt{R}_1$ & 6 & 0.329 & 0.145 & 0.091 & 0.079 & 0.356 \\
$\mathtt{R}_5\mathtt{R}_2$ & 5 & 0.273 & 0.383 & 0.105 & 0.154 & 0.086 \\
$\mathtt{R}_5\mathtt{R}_3$ & 3 & 0.143 & 0.125 & 0.280 & 0.167 & 0.286 \\
$\mathtt{R}_5\mathtt{R}_5$ & 4 & 0.351 & 0.190 & 0.112 & 0.121 & 0.225 \\
$\mathtt{R}_5\mathtt{R}_5\mathtt{R}_4$ & 3 & 0.000 & 0.317 & 0.000 & 0.583 & 0.100 \\
\hline
\end{tabular}
\end{table}

As a comparison reference for the Extreme-regime context tree, analogous NYA context trees for the High (2022) and Normal (2005) regimes are reported in Figs.~\ref{fig:vlmc-case2} and~\ref{fig:vlmc-case3} in the Appendix. The corresponding aggregated context tables for developed stock market indices are provided in Tables~\ref{tab:developed-high} and~\ref{tab:developed-normal}. For developing stock market indices, the aggregated context tables for the Extreme, High, and Normal regimes are reported in Tables~\ref{tab:developing-extreme}, \ref{tab:developing-high}, and~\ref{tab:developing-normal}, respectively, in the Appendix.


We compare the unconditional probabilities of $\mathtt{R}_1, \mathtt{R}_2, \mathtt{R}_3, \mathtt{R}_4 \hspace{2pt} \text{and}\hspace{2pt} \mathtt{R}_5$, the tail ratio \\ $\Bigg(\displaystyle\frac{P(\mathtt{R}_1)+P(\mathtt{R}_5)}{P(\mathtt{R}_2)+P(\mathtt{R}_3)+P(\mathtt{R}_4)}\Bigg)$, and Shannon entropy $\Big(\displaystyle-\sum_iP(\mathtt{R}_i)\log_2P(\mathtt{R}_i)\Big)$ across regimes, as shown in Table~\ref{tab:unconditional-probabilities}. In Extreme regimes, both markets show elevated tail risks. For developed markets, $\mathtt{R}_1=31.15\%$ and $\mathtt{R}_5=27.13\%$, while for developing markets, $\mathtt{R}_1=29.91\%$ and $\mathtt{R}_5=23.99\%$. These are accompanied by suppressed middle states. In developed markets, $\mathtt{R}_2=15.91\%$, $\mathtt{R}_3=12.14\%$, and $\mathtt{R}_4=13.67\%$, while in developing markets, $\mathtt{R}_2=16.84\%$, $\mathtt{R}_3=14.22\%$, and $\mathtt{R}_4=15.02\%$. This indicates high susceptibility to large price swings. This tail risk decreases as the regime changes from Extreme to High to then Normal, signaling market stability where extreme outcomes become less frequent.

Across both markets, left-tail risk $\mathtt{R}_1$ consistently exceeds right-tail risk $\mathtt{R}_5$, signaling a persistent downside-risk asymmetry. The magnitude of that asymmetry, however, varies by regime. In Extreme regimes, developing markets show greater downside risk, with a 5.92\% spread compared to 4.02\% in developed markets, suggesting that panic-driven accelerated sell-offs are more prevalent in developing economies, in line with earlier findings~\cite{li2009tail,pereda2025systemic}. In Normal regimes too, the asymmetry remains pronounced, with a 2.33\% spread compared to 0.15\% in developed markets. This may indicate the fragmented nature of developing markets where there are information delays and liquidity constraints, leading to amplification of negative shocks even when volatility is low, as documented in~\cite{lesmond2005liquidity}. However, the pattern reverses in High regimes. Developed markets show the greater spread, at 5.30\% versus 2.86\% in developing markets. One likely reason for this is that institutional investors hedge heavily, and their protective trades can add extra downside risk, as documented empirically in these studies~\cite{garleanu2008demand,coval2007asset}.

The tail ratio also decreases sharply from Extreme to Normal regimes in both markets, but the attenuation is more pronounced in developed markets, falling from 1.3969 to 0.2552, an 81.7\% decrease, than in developing markets, falling from 1.1697 to 0.3457, a 70.5\% decrease. Developing markets exhibit a 35\% higher tail ratio during Normal periods, quantifying their persistent tail risk exposure. Entropy peaks in High regimes, with 2.3071 in developed markets and 2.3059 in developing markets, indicating maximum unpredictability during High regimes, and declines in Extreme and Normal regimes. Thus, Extreme regimes concentrate tail risks, High regimes maximize uncertainty, and Normal regimes minimize extremes, with developing markets demonstrating systematically higher residual tail risk.

\begin{table}[H]
\centering
\footnotesize
\setlength{\tabcolsep}{7pt}
\renewcommand{\arraystretch}{1.3}
\caption{Unconditional Probabilities with Tail Ratio and Shannon Entropy}
\label{tab:unconditional-probabilities}
\scalebox{1.1}{%
  \begin{tabular}{|c|c|r|r|r|r|r|c|c|}
    \hline
    \multirow{2}{*}{\makecell{Market}}
      & \multirow{2}{*}{\makecell{Regime}}
        & \multicolumn{5}{c|}{Unconditional probabilities}
          & \multirow{2}{*}{Tail ratio}
            & \multirow{2}{*}{\makecell{Shannon\\entropy}} \\
    \cline{3-7}
    &  & P($\mathtt{R}_1$) & P($\mathtt{R}_2$) & P($\mathtt{R}_3$) & P($\mathtt{R}_4$) & P($\mathtt{R}_5$) &  &  \\
    \hline
    \multirow{3}{*}{Developed}
      & Extreme  & 0.312 & 0.159 & 0.121 & 0.137 & 0.271 & 1.397 & 2.219 \\
    \cline{2-9}
      & High     & 0.251 & 0.198 & 0.162 & 0.191 & 0.198 & 0.816 & 2.307 \\
    \cline{2-9}
      & Normal   & 0.102 & 0.283 & 0.278 & 0.235 & 0.101 & 0.255 & 2.191 \\
    \hline
    \multirow{3}{*}{Developing}
      & Extreme  & 0.299 & 0.168 & 0.142 & 0.150 & 0.240 & 1.170 & 2.259 \\
    \cline{2-9}
      & High     & 0.248 & 0.192 & 0.174 & 0.167 & 0.219 & 0.877 & 2.306 \\
    \cline{2-9}
      & Normal   & 0.140 & 0.251 & 0.276 & 0.217 & 0.117 & 0.346 & 2.250 \\
    \hline
  \end{tabular}
}
\end{table}

Following the unconditional probability analysis, we examine conditional probabilities at different orders $k$. For $k=1$, where the next return depends only on the most recent return state, we summarize the context-tree transitions using two metrics defined in Section~\ref{subsec:vlmc_metrics}: self-persistence $\mathtt{M}$, capturing one-step repetition $\mathtt{R}_i\to\mathtt{R}_i$, and reversal intensity $\mathtt{V}$, capturing paired flips. Table~\ref{tab:order1_metrics} reports these $k=1$ metrics across regimes and markets.

In developed markets, tail persistence peaks in the Extreme regime, with $\mathtt{M}_1=0.303$ and $\mathtt{M}_5=0.263$, and then declines through High to Normal, where $\mathtt{M}_1=0.278$ and $\mathtt{M}_5=0.202$ in High and $\mathtt{M}_1=0.182$ and $\mathtt{M}_5=0.105$ in Normal. Mid-range persistence across $\mathtt{M}_2$--$\mathtt{M}_4$ rises sharply, most notably $\mathtt{M}_3$, which increases from $0.147$ in Extreme to $0.270$ in Normal. The same stabilization pattern is also observed in developing markets as regimes transition from Extreme to High to Normal. However, developing markets show higher downside persistence across all regimes, with $\mathtt{M}_1=0.336$ in Extreme, $\mathtt{M}_1=0.295$ in High, and $\mathtt{M}_1=0.254$ in Normal, each exceeding the corresponding developed-market values. This confirms chronic downside stickiness, in which negative states persist longer in developing markets.

For reversal intensity, developed markets show the strongest tail-to-tail flipping in the Extreme regime, with $\mathtt{V}_1=0.329$, which then declines through High to Normal, where $\mathtt{V}_1=0.232$ in High and $\mathtt{V}_1=0.106$ in Normal. In contrast, moderate reversals strengthen as regimes stabilize, with $\mathtt{V}_2=0.199$ in Extreme, $\mathtt{V}_2=0.214$ in High, and $\mathtt{V}_2=0.283$ in Normal. Developing markets exhibit the same qualitative pattern but with distinct magnitudes. In the Extreme regime, tail flips are weaker in developing markets, with $\mathtt{V}_1=0.267$ versus $0.329$ in developed markets, indicating that large tail-to-tail corrections occur less frequently in extremely volatile periods for developing indices. In Normal regimes, the pattern reverses: developing markets show stronger tail reversals, with $\mathtt{V}_1=0.132$ versus $0.106$, consistent with greater sensitivity to liquidity frictions and information asymmetry prevalent in developing economies.

Overall, the $k=1$ patterns indicate progressive stabilization from Extreme to Normal regimes in both markets. However, developing markets exhibit persistent downside stickiness, with higher downside persistence across all regimes, indicating more prolonged vulnerability to negative shocks. They also show asymmetric tail reversals: tail-to-tail flipping is weaker in Extreme regimes yet comparatively stronger in Normal regimes, relative to developed markets. This reflects structural weaknesses such as thinner liquidity and information delays in developing markets. Market participants must therefore recognize these vulnerabilities and strategies should be adapted accordingly --- expect slower price reversals during extreme volatility periods, prepare for extended downturns requiring greater patience, and tailor risk management to address persistent negative states.

\newcolumntype{C}[1]{>{\centering\arraybackslash}p{#1}} 

\begin{table}[H]
\centering
\scriptsize
\setlength{\tabcolsep}{7pt}
\renewcommand{\arraystretch}{1.3}
\caption{Context‐tree metrics for order $k=1$.}
\label{tab:order1_metrics}
\scalebox{1.1}{%
  \begin{tabular}{|c|c|r|r|r|r|r|C{1.5cm}|C{1.5cm}|}
    \hline
    \multirow{2}{*}{\makecell{Market}}
      & \multirow{2}{*}{\makecell{Regime}}
        & \multicolumn{5}{c|}{Self‐persistence ($\mathtt{M}$)}
          & \multicolumn{2}{c|}{Reversal intensity ($\mathtt{V}$)} \\
    \cline{3-9}
    & & $\mathtt{M}_1$ & $\mathtt{M}_2$ & $\mathtt{M}_3$ & $\mathtt{M}_4$ & $\mathtt{M}_5$ & $\mathtt{V}_1$ & $\mathtt{V}_2$ \\
    \hline
    \multirow{3}{*}{Developed}
      & Extreme & 0.303 & 0.174 & 0.147 & 0.115 & 0.263 & 0.329 & 0.199 \\
    \cline{2-9}
      & High    & 0.278 & 0.170 & 0.177 & 0.185 & 0.202 & 0.232 & 0.214 \\
    \cline{2-9}
      & Normal  & 0.182 & 0.260 & 0.270 & 0.210 & 0.105 & 0.106 & 0.283 \\
    \hline
    \multirow{3}{*}{Developing}
      & Extreme & 0.336 & 0.188 & 0.168 & 0.171 & 0.253 & 0.267 & 0.182 \\
    \cline{2-9}
      & High    & 0.295 & 0.185 & 0.156 & 0.179 & 0.225 & 0.230 & 0.176 \\
    \cline{2-9}
      & Normal  & 0.254 & 0.242 & 0.281 & 0.180 & 0.145 & 0.132 & 0.267 \\
    \hline
  \end{tabular}
}
\end{table}

Extending the conditional analysis beyond $k=1$, we examine higher-order dependence at $k=2$ and $k=3$, where the next return state depends on the previous two or three states. Table~\ref{tab:context_tree_metrics} reports the corresponding context-tree metrics defined in Section~\ref{subsec:vlmc_metrics}: Continuation ($\mathtt{C}_k$), Exhaustion ($\mathtt{E}_k$), Zigzag alternation $\mathtt{Z}_k$, and Burst-from-calm ($\mathtt{B}_k$).

For $k=2$ in developed markets, continuation is highest in the Extreme regime with $\mathtt{C}_2=0.063$, drops in the High regime to $\mathtt{C}_2=0.033$, and rebounds in the Normal regime to $\mathtt{C}_2=0.058$. This pattern indicates that two-day runs are most persistent under stress, weaken during High regimes, and partially re-emerge in Normal conditions, which may suggest restored confidence where orderly price discovery allows trend-following behavior to regain momentum. Exhaustion declines with stabilization, from $\mathtt{E}_2=0.379$ in the Extreme regime to $\mathtt{E}_2=0.271$ in the Normal regime, consistent with fewer sharp tail-to-tail corrections outside the most turbulent periods. In other words, the abrupt reversals associated with panic-driven sell-offs and subsequent rebound buying become less dominant as markets stabilize. Zigzag alternation also drops strongly, from $\mathtt{Z}_2=0.362$ in the Extreme regime to $\mathtt{Z}_2=0.100$ in the Normal regime, showing that tail-to-tail whipsaw dynamics become substantially less pronounced as volatility conditions normalize. In contrast, burst-from-calm strengthens outside Extreme regimes, with $\mathtt{B}_2=0.117$ in the Extreme regime and $\mathtt{B}_2=0.140$ in the Normal regime, indicating that tail moves can still arise from calm sequences even when overall volatility is low.

Developing markets exhibit both similarities to, and clear departures from, developed markets at $k=2$. In common with developed markets, continuation is weakest in the High regime, with $\mathtt{C}_2=0.031$, and burst-from-calm is larger outside the Extreme regime, rising from $\mathtt{B}_2=0.106$ in the Extreme regime to $\mathtt{B}_2=0.142$ in the Normal regime. These patterns indicate that, in both developed and developing markets, two-day run continuation weakens in the intermediate High regime and tail events can still emerge from calm sequences as regimes move toward Normal conditions. The differences are most apparent in the Normal regime. Continuation becomes substantially stronger in developing markets, with $\mathtt{C}_2=0.110$ versus $0.058$ in developed markets, indicating more pronounced two-day run persistence even under minimal-volatility conditions. Exhaustion is also markedly higher, with $\mathtt{E}_2=0.401$ versus $0.271$ in developed markets, suggesting that tail-to-tail switching after extreme runs remains more prevalent in developing markets during stable periods, consistent with shallower liquidity conditions. Zigzag alternation likewise remains elevated, with $\mathtt{Z}_2=0.225$ in the Normal regime compared with $0.100$ in developed markets, pointing to more persistent tail switching and noisier short-horizon dynamics. Overall, these contrasts indicate that developing markets retain stronger higher-order tail dependence and greater residual fragility even when regimes are classified as Normal.

\newcolumntype{C}[1]{>{\centering\arraybackslash}p{#1}}
\begin{table}[H]
\centering
\scriptsize
\setlength{\tabcolsep}{7pt}
\renewcommand{\arraystretch}{1.4}
\caption{Context‐tree metrics for order $k=2,3$.}
\label{tab:context_tree_metrics}
\scalebox{1.1}{%
  \begin{tabular}{|c|c|r|r|r|r|r|r|r|r|}
    \hline
    \multirow{2}{*}{\makecell{Market}}
      & \multirow{2}{*}{\makecell{Regime}}
        & \multicolumn{4}{c|}{$k=2$}
        & \multicolumn{4}{c|}{$k=3$} \\
    \cline{3-10}
    & & $\mathtt{C}_2$ & $\mathtt{E}_2$ & $\mathtt{Z}_2$ & $\mathtt{B}_2$ & $\mathtt{C}_3$ & $\mathtt{E}_3$ & $\mathtt{Z}_3$ & $\mathtt{B}_3$ \\
    \hline
    \multirow{3}{*}{Developed}
      & Extreme 
        & 0.063 & 0.379 & 0.362 & 0.117
        & 0.018 & 0.490 & 0.233 & $\ll0.001$ \\
    \cline{2-10}
      & High    
        & 0.033 & 0.250 & 0.296 & 0.139
        & $\ll0.001$ & $\ll0.001$ & 0.083 & $\ll0.001$ \\
    \cline{2-10}
      & Normal  
        & 0.058 & 0.271 & 0.100 & 0.140
        & $\ll0.001$ & $\ll0.001$ & $\ll0.001$ & 0.259 \\
    \hline
    \multirow{3}{*}{Developing}
      & Extreme 
        & 0.072 & 0.275 & 0.298 & 0.106
        & 0.008 & 0.250 & 0.125 & 0.060 \\
    \cline{2-10}
      & High    
        & 0.031 & 0.285 & 0.212 & 0.079
        & $\ll0.001$ & 0.196 & $\ll0.001$ & 0.029 \\
    \cline{2-10}
      & Normal  
        & 0.110 & 0.401 & 0.225 & 0.142
        & 0.033 & 0.250 & $\ll0.001$ & 0.124 \\
    \hline
  \end{tabular}
}
\end{table}

For $k=3$, developed markets show three-day dependence primarily in the Extreme regime. Continuation is detectable with $\mathtt{C}_3=0.018$, and exhaustion is elevated with $\mathtt{E}_3=0.490$, while zigzag alternation remains sizeable with $\mathtt{Z}_3=0.233$. Together, these results indicate that three-day patterns are most evident under stress, where multi-day sequences can persist and then flip sharply. In the High regime, three-day dependence largely vanishes, with only a modest zigzag signal $\mathtt{Z}_3=0.083$. In the Normal regime, continuation and tail switching are negligible, yet burst-from-calm becomes dominant with $\mathtt{B}_3=0.259$, indicating that long calm sequences can still mask the risk of a sudden extreme move.

Developing markets retain more three-day structure. In the Extreme regime, $\mathtt{C}_3=0.008$, $\mathtt{E}_3=0.250$, $\mathtt{Z}_3=0.125$, and $\mathtt{B}_3=0.060$ are all present, mirroring the qualitative profile of developed markets but at lower magnitudes. In the High regime, exhaustion remains detectable with $\mathtt{E}_3=0.196$ and burst-from-calm persists with $\mathtt{B}_3=0.029$, indicating that higher-order tail switching does not fully disappear. In the Normal regime, continuation reappears with $\mathtt{C}_3=0.033$, exhaustion remains at $\mathtt{E}_3=0.250$, and burst-from-calm rises to $\mathtt{B}_3=0.124$, implying that multi-day persistence and tail switching can coexist with non-negligible burst risk even under minimal-volatility conditions. Overall, relative to developed markets, developing markets exhibit more persistent higher-order dependence and a stronger tendency for tail-related dynamics to remain active in Normal regimes.

\section{Discussion}
\label{sec:Disc}

Financial market conditions are rarely uniform over time. Periods of routine trading with moderate fluctuations alternate with episodes of heightened stress marked by sharp price swings. Classical regime frameworks describe this alternation as shifts between low-volatility and high-volatility or calm and stressed market states~\cite{hamilton1989new,ang2012regime,guidolin2007asset}. This alternation motivates the need for an indicator that can detect regime shifts reliably in financial data. Standard volatility measures, including rolling standard deviations and GARCH-type conditional variances, provide useful summaries of volatility levels~\cite{engle1982autoregressive,bollerslev1986generalized}. However, they are backward-looking by construction. As a result, they can adjust slowly when volatility changes abruptly. Structural breaks may then appear as spurious persistence in GARCH dynamics~\cite{lamoureux1990persistence}. Predictive performance can also deteriorate when the data-generating process shifts sharply during crises~\cite{hillebrand2010benefits}.

To address these limitations, we operationalize regime identification using instantaneous energy from the Hilbert--Huang Transform. In addition, GARCH-based indicators require parametric specifications for volatility evolution and innovation distributions, whereas the HHT-based energy measure is obtained through a data-driven decomposition and remains informative when volatility is shaped by liquidity stress and feedback mechanisms~\cite{brunnermeier2009deciphering,shleifer2011fire}. We use thresholds at $\mu+\sigma$ and $\mu+6\sigma$, where $\mu$ and $\sigma$ denote the mean and standard deviation of instantaneous energy. The $\mu+\sigma$ cutoff marks sustained departures from baseline energy levels and separates calm conditions from periods of elevated volatility, while $\mu+6\sigma$ is deliberately conservative and isolates only the most extreme energy realizations. When $E(t)>\mu+6\sigma$, oscillatory activity becomes sharply amplified across intrinsic time scales, consistent with severe stress dynamics such as liquidity withdrawal, widening bid--ask spreads, forced liquidations, and adverse feedback loops~\cite{brunnermeier2009deciphering,shleifer2011fire}. These cutoffs act as scale-normalized separators in a heavy-tailed financial system rather than Gaussian tail-probability statements~\cite{cont2001empirical,embrechts2013modelling}. The resulting regimes align with classical interpretations. Normal corresponds to low-volatility liquid conditions. High corresponds to sustained elevated volatility in functioning markets. Extreme corresponds to rare stress episodes with severe dislocations. Robustness is assessed via sensitivity analysis under alternative threshold specifications, as detailed in~\ref{appendix:sensitivity}. The findings of the sensitivity analysis reported in Figshare at~\href{https://doi.org/10.6084/m9.figshare.30982552}{10.6084/m9.figshare.30982552} show that qualitative conclusions are unchanged under reasonable variations.


Beyond regime detection, HHSA is introduced to complement the HHT-based segmentation and to justify the subsequent VLMC analysis by providing an independent, scale-resolved description of within-regime volatility organization. While instantaneous energy identifies when markets enter higher-activity states, it does not describe how volatility is structured across time scales inside each state. HHSA addresses this by resolving cross-frequency dynamics and extracting carrier frequencies $\omega_c$ and amplitude-modulation frequencies $\omega_{am}$~\cite{huang2016holo,nguyen2019unraveling}. This regime-level volatility profiling provides empirical evidence that the regimes identified by instantaneous energy differ not only in level but also in internal volatility structure. Consistent with this interpretation, amplitude-modulation energy declines by a factor of $15$ from Extreme to Normal regimes in developed markets, compared to $\approx 7$-fold in developing markets, indicating systematic differences in how markets dissipate stress across regimes. Taken together, HHT provides a time-localized basis for regime segmentation, HHSA quantifies within-regime volatility structure, and VLMC is then applied to examine how return-category transitions evolve conditionally on these empirically distinct regimes.


\section{Conclusions}
\label{sec:Conc}

Financial markets exhibit abrupt transitions between tranquil and stressed periods, and return dynamics can change across such regimes. Identifying these regimes and quantifying how return dynamics differ across them is important for risk management and portfolio allocation, especially when tail events cluster and volatility conditions vary by market maturity. This study examines regime-dependent return dynamics in developed and developing equity indices by combining Hilbert--Huang based regime identification and profiling with a Variable-Length Markov Chain analysis of categorized returns. 

Market regimes are first identified using Empirical Mode Decomposition based Hilbert--Huang Transform. Following regime identification, we profile each regime using Holo--Hilbert Spectral Analysis. The profiles show systematic regime-dependent shifts in price dynamics and volatility behavior that differ fundamentally between developed and developing markets, thus providing empirical support for examining return-state transitions separately within each regime. To examine regime-dependent return dynamics, daily index returns are categorized into discrete states and analyzed using variable-length Markov chains. The unconditional probabilities reveal that while the prevalence of extreme returns recedes as regimes stabilize, a persistent downside asymmetry remains across all regimes. The reduction in extreme outcomes is significantly more pronounced in developed markets. In contrast, developing markets retain persistent tail exposure even under minimal-volatility conditions. Furthermore, market unpredictability is observed to peak during moderate volatility periods. Conditional transition dynamics indicate progressive stabilization from Extreme to Normal regimes in both markets, though developing markets retain clear downside persistence. Reversal intensity is observed to be regime-dependent, characterized by weaker tail-to-tail transitions in Extreme regimes and stronger tail reversals in Normal regimes. Higher-order dependence further differentiates the two market groups. Developed markets exhibit a pronounced reduction in tail alternation as regimes normalize, consistent with the efficient dissipation of whipsaw-type dynamics. In contrast, developing markets maintain elevated exhaustion and zigzag alternation, indicating that higher-order tail dependence remains active even under low-volatility conditions. 


Overall, the findings show that regime dependence is present in both developed and developing markets, but the nature of stabilization differs materially with market maturity. Developed markets transition toward more ordered conditional dynamics and weaker tail-dependent structure as regimes normalize. Developing markets retain residual fragility, with persistent higher-order tail dynamics and non-negligible burst risk even in Normal regimes. These results imply that developing markets may require targeted safeguards and risk controls not only during crises but also during stable periods, reflecting structural frictions such as thinner liquidity and slower information incorporation. The findings are consistent with prior evidence on tail-risk asymmetries and crisis amplification, liquidity-friction effects across market maturity, and volatility–return dynamics shaped by investor demand and intermediary constraints \cite{li2009tail,pereda2025systemic,lesmond2005liquidity,garleanu2008demand,coval2007asset}. 

A limitation of the present analysis is that the estimated context trees condition only on past return states and do not incorporate observable external drivers that may influence transitions. Future work can extend this framework by incorporating exogenous drivers through covariate-dependent variable-length Markov models, allowing transition probabilities and context selection to vary with economic conditions. Suitable covariates include realized volatility and volume, liquidity proxies such as bid--ask spreads, policy and macro indicators, and global risk indicators, thereby linking regime-dependent dynamics more directly to measurable market drivers.

\section*{Acknowledgements}
The authors, S. R. Luwang, K. Mukhia, and B. N. Sharma, would like to thank the National Institute of Technology Sikkim, for allocating doctoral research fellowships.

\section*{Data and Code Availability Statement}
The data that support the findings of this study are publicly available from \href{https://finance.yahoo.com/}{Yahoo Finance}. The code used for regime identification and profiling using HHT and HHSA is provided in the form of Jupyter notebooks in Figshare at \href{https://doi.org/10.6084/m9.figshare.30982552}{10.6084/m9.figshare.30982552}. The code used for intra-regime return-dynamics analysis via VLMC context trees is also provided in Figshare at \href{https://doi.org/10.6084/m9.figshare.30982552}{10.6084/m9.figshare.30982552} and is implemented using the \textit{mixvlmc} package in R.

\begingroup
\small                
\setlength{\bibsep}{2pt}      
\bibliographystyle{elsarticle-num}
\bibliography{References}

@article{hamilton1990analysis,
  title={Analysis of time series subject to changes in regime},
  author={Hamilton, James D},
  journal={Journal of econometrics},
  volume={45},
  number={1-2},
  pages={39--70},
  year={1990},
  publisher={Elsevier}
}

@article{ang2012regime,
  title={Regime changes and financial markets},
  author={Ang, Andrew and Timmermann, Allan},
  journal={Annu. Rev. Financ. Econ.},
  volume={4},
  number={1},
  pages={313--337},
  year={2012},
  publisher={Annual Reviews}
}

@article{engle2004risk,
  title={Risk and volatility: Econometric models and financial practice},
  author={Engle, Robert},
  journal={American economic review},
  volume={94},
  number={3},
  pages={405--420},
  year={2004},
  publisher={American Economic Association}
}

@article{longin1996asymptotic,
  title={The asymptotic distribution of extreme stock market returns},
  author={Longin, Fran{\c{c}}ois M},
  journal={Journal of business},
  pages={383--408},
  year={1996},
  publisher={JSTOR}
}

@article{rey2014detection,
  title={Detection of high and low states in stock market returns with MCMC method in a Markov switching model},
  author={Rey, Cl{\'e}ment and Rey, Serge and Viala, Jean-Renaud},
  journal={Economic Modelling},
  volume={41},
  pages={145--155},
  year={2014},
  publisher={Elsevier}
}

@article{bensaida2015frequency,
  title={The frequency of regime switching in financial market volatility},
  author={BenSa{\"\i}da, Ahmed},
  journal={Journal of Empirical Finance},
  volume={32},
  pages={63--79},
  year={2015},
  publisher={Elsevier}
}

@article{mandelbrot1963variation,
  title={The variation of certain speculative prices},
  author={Mandelbrot, Benoit and others},
  journal={Journal of business},
  volume={36},
  number={4},
  pages={394},
  year={1963},
  publisher={Springer}
}

@article{schwert1989does,
  title={Why does stock market volatility change over time?},
  author={Schwert, G William},
  journal={The journal of finance},
  volume={44},
  number={5},
  pages={1115--1153},
  year={1989},
  publisher={Wiley Online Library}
}

@article{andersen1998answering,
  title={Answering the skeptics: Yes, standard volatility models do provide accurate forecasts},
  author={Andersen, Torben G and Bollerslev, Tim},
  journal={International economic review},
  pages={885--905},
  year={1998},
  publisher={JSTOR}
}

@article{mikosch2004nonstationarities,
  title={Nonstationarities in financial time series, the long-range dependence, and the IGARCH effects},
  author={Mikosch, Thomas and St{\u{a}}ric{\u{a}}, C{\u{a}}t{\u{a}}lin},
  journal={Review of Economics and Statistics},
  volume={86},
  number={1},
  pages={378--390},
  year={2004},
  publisher={MIT Press 238 Main St., Suite 500, Cambridge, MA 02142-1046, USA journals~…}
}

@book{gatheral2011volatility,
  title={The volatility surface: a practitioner's guide},
  author={Gatheral, Jim},
  year={2011},
  publisher={John Wiley \& Sons}
}

@article{kritzman2012regime,
  title={Regime shifts: Implications for dynamic strategies (corrected)},
  author={Kritzman, Mark and Page, Sebastien and Turkington, David},
  journal={Financial Analysts Journal},
  volume={68},
  number={3},
  pages={22--39},
  year={2012},
  publisher={Taylor \& Francis}
}

@article{lux2000volatility,
  title={Volatility clustering in financial markets: a microsimulation of interacting agents},
  author={Lux, Thomas and Marchesi, Michele},
  journal={International journal of theoretical and applied finance},
  volume={3},
  number={04},
  pages={675--702},
  year={2000},
  publisher={World Scientific}
}

@incollection{engle2007good,
  title={What good is a volatility model?},
  author={Engle, Robert F and Patton, Andrew J},
  booktitle={Forecasting volatility in the financial markets},
  pages={47--63},
  year={2007},
  publisher={Elsevier}
}

@article{guidolin2007asset,
  title={Asset allocation under multivariate regime switching},
  author={Guidolin, Massimo and Timmermann, Allan},
  journal={Journal of Economic Dynamics and Control},
  volume={31},
  number={11},
  pages={3503--3544},
  year={2007},
  publisher={Elsevier}
}

@article{nystrup2015regime,
  title={Regime-based versus static asset allocation: Letting the data speak},
  author={Nystrup, Peter and Hansen, Bo William and Madsen, Henrik and Lindstr{\"o}m, Erik},
  journal={Journal of Portfolio Management},
  volume={42},
  number={1},
  pages={103},
  year={2015},
  publisher={Pageant Media}
}

@article{d2012weighted,
  title={Weighted-indexed semi-Markov models for modeling financial returns},
  author={D’Amico, Guglielmo and Petroni, Filippo},
  journal={Journal of statistical mechanics: theory and experiment},
  volume={2012},
  number={07},
  pages={P07015},
  year={2012},
  publisher={IOP Publishing}
}

@article{rai2023detection,
  title={Detection and forecasting of extreme events in stock price triggered by fundamental, technical, and external factors},
  author={Rai, Anish and Luwang, Salam Rabindrajit and Nurujjaman, Md and Hens, Chittaranjan and Kuila, Pratyay and Debnath, Kanish},
  journal={Chaos, Solitons \& Fractals},
  volume={173},
  pages={113716},
  year={2023},
  publisher={Elsevier}
}

@article{mahata2021characteristics,
  title={Characteristics of 2020 stock market crash: The COVID-19 induced extreme event},
  author={Mahata, Ajit and Rai, Anish and Nurujjaman, Md and Prakash, Om and Prasad Bal, Debi},
  journal={Chaos: An Interdisciplinary Journal of Nonlinear Science},
  volume={31},
  number={5},
  year={2021},
  publisher={AIP Publishing}
}

@article{ang2007stock,
  title={Stock return predictability: Is it there?},
  author={Ang, Andrew and Bekaert, Geert},
  journal={The Review of Financial Studies},
  volume={20},
  number={3},
  pages={651--707},
  year={2007},
  publisher={Oxford University Press}
}

@article{flannery2002macroeconomic,
  title={Macroeconomic factors do influence aggregate stock returns},
  author={Flannery, Mark J and Protopapadakis, Aris A},
  journal={The review of financial studies},
  volume={15},
  number={3},
  pages={751--782},
  year={2002},
  publisher={Oxford University Press}
}

@article{avramov2006predicting,
  title={Predicting stock returns},
  author={Avramov, Doron and Chordia, Tarun},
  journal={Journal of Financial Economics},
  volume={82},
  number={2},
  pages={387--415},
  year={2006},
  publisher={Elsevier}
}

@article{marquering2004economic,
  title={The economic value of predicting stock index returns and volatility},
  author={Marquering, Wessel and Verbeek, Marno},
  journal={Journal of Financial and Quantitative Analysis},
  volume={39},
  number={2},
  pages={407--429},
  year={2004},
  publisher={Cambridge University Press}
}

@article{zanin2022variable,
  title={Variable length Markov chain with exogenous covariates},
  author={Zanin Zambom, Adriano and Kim, Seonjin and Lopes Garcia, Nancy},
  journal={Journal of Time Series Analysis},
  volume={43},
  number={2},
  pages={312--328},
  year={2022},
  publisher={Wiley Online Library}
}

@article{reboredo2010nonlinear,
  title={Nonlinear effects of oil shocks on stock returns: a Markov-switching approach},
  author={Reboredo, Juan C},
  journal={Applied Economics},
  volume={42},
  number={29},
  pages={3735--3744},
  year={2010},
  publisher={Taylor \& Francis}
}

@article{huang2017applying,
  title={Applying a Markov chain for the stock pricing of a novel forecasting model},
  author={Huang, Jui-Chieh and Huang, Wen-Tso and Chu, Pei-Tzu and Lee, Wen-Yi and Pai, Hsin-Ping and Chuang, Chih-Chen and Wu, Ya-Wen},
  journal={Communications in Statistics-theory and Methods},
  volume={46},
  number={9},
  pages={4388--4402},
  year={2017},
  publisher={Taylor \& Francis}
}

@article{buhlmann1999variable,
  title={Variable length Markov chains},
  author={B{\"u}hlmann, Peter and Wyner, Abraham J},
  journal={The Annals of Statistics},
  volume={27},
  number={2},
  pages={480--513},
  year={1999},
  publisher={Institute of Mathematical Statistics}
}

@article{chang2009macroeconomic,
  title={Do macroeconomic variables have regime-dependent effects on stock return dynamics? Evidence from the Markov regime switching model},
  author={Chang, Kuang-Liang},
  journal={Economic Modelling},
  volume={26},
  number={6},
  pages={1283--1299},
  year={2009},
  publisher={Elsevier}
}

@article{rabindrajit2024high,
  title={High-frequency stock market order transitions during the US--China trade war 2018: A discrete-time Markov chain analysis},
  author={Rabindrajit Luwang, Salam and Rai, Anish and Nurujjaman, Md and Prakash, Om and Hens, Chittaranjan},
  journal={Chaos: An Interdisciplinary Journal of Nonlinear Science},
  volume={34},
  number={1},
  year={2024},
  publisher={AIP Publishing}
}

@article{garleanu2008demand,
  title={Demand-based option pricing},
  author={Garleanu, Nicolae and Pedersen, Lasse Heje and Poteshman, Allen M},
  journal={The Review of Financial Studies},
  volume={22},
  number={10},
  pages={4259--4299},
  year={2008},
  publisher={Society for Financial Studies}
}

@article{coval2007asset,
  title={Asset fire sales (and purchases) in equity markets},
  author={Coval, Joshua and Stafford, Erik},
  journal={Journal of Financial Economics},
  volume={86},
  number={2},
  pages={479--512},
  year={2007},
  publisher={Elsevier}
}

@article{hillebrand2010benefits,
  title={The benefits of bagging for forecast models of realized volatility},
  author={Hillebrand, Eric and Medeiros, Marcelo C},
  journal={Econometric Reviews},
  volume={29},
  number={5-6},
  pages={571--593},
  year={2010},
  publisher={Taylor \& Francis}
}

@article{broock1996test,
  title={A test for independence based on the correlation dimension},
  author={Broock, William A and Scheinkman, Jos{\'e} Alexandre and Dechert, W Davis and LeBaron, Blake},
  journal={Econometric reviews},
  volume={15},
  number={3},
  pages={197--235},
  year={1996},
  publisher={Taylor \& Francis}
}

@article{hsieh1991chaos,
  title={Chaos and nonlinear dynamics: application to financial markets},
  author={Hsieh, David A},
  journal={The journal of finance},
  volume={46},
  number={5},
  pages={1839--1877},
  year={1991},
  publisher={Wiley Online Library}
}

@book{brock1991nonlinear,
  title={Nonlinear dynamics, chaos, and instability: statistical theory and economic evidence},
  author={Brock, William A and Hsieh, David Arthur and LeBaron, Blake Dean},
  year={1991},
  publisher={MIT press}
}

@article{huang2016holo,
  title={On holo-Hilbert spectral analysis: a full informational spectral representation for nonlinear and non-stationary data},
  author={Huang, Norden E and Hu, Kun and Yang, Albert CC and Chang, Hsing-Chih and Jia, Deng and Liang, Wei-Kuang and Yeh, Jia Rong and Kao, Chu-Lan and Juan, Chi-Hung and Peng, Chung Kang and others},
  journal={Philosophical Transactions of the Royal Society A: Mathematical, Physical and Engineering Sciences},
  volume={374},
  number={2065},
  pages={20150206},
  year={2016},
  publisher={The Royal Society Publishing}
}

@article{nguyen2019unraveling,
  title={Unraveling nonlinear electrophysiologic processes in the human visual system with full dimension spectral analysis},
  author={Nguyen, Kien Trong and Liang, Wei-Kuang and Lee, Victor and Chang, Wen-Sheng and Muggleton, Neil G and Yeh, Jia-Rong and Huang, Norden E and Juan, Chi-Hung},
  journal={Scientific reports},
  volume={9},
  number={1},
  pages={16919},
  year={2019},
  publisher={Nature Publishing Group UK London}
}

@article{huang2009instantaneous,
  title={On instantaneous frequency},
  author={Huang, Norden E and Wu, Zhaohua and Long, Steven R and Arnold, Kenneth C and Chen, Xianyao and Blank, Karin},
  journal={Advances in adaptive data analysis},
  volume={1},
  number={02},
  pages={177--229},
  year={2009},
  publisher={World Scientific}
}

@article{huang2013uniqueness,
  title={The uniqueness of the instantaneous frequency based on intrinsic mode function},
  author={Huang, Norden E and YOUNG, VINCENT and LO, MENTZUNG and WANG, YUNG HUNG and Peng, Chung-Kang and Chen, Xianyao and Wang, Gang and Deng, Jia and Wu, Zhaohua},
  journal={Advances in adaptive data analysis},
  volume={5},
  number={03},
  pages={1350011},
  year={2013},
  publisher={World Scientific}
}

@inproceedings{takens2006detecting,
  title={Detecting strange attractors in turbulence},
  author={Takens, Floris},
  booktitle={Dynamical Systems and Turbulence, Warwick 1980: proceedings of a symposium held at the University of Warwick 1979/80},
  pages={366--381},
  year={2006},
  organization={Springer}
}

@article{grassberger1983measuring,
  title={Measuring the strangeness of strange attractors},
  author={Grassberger, Peter and Procaccia, Itamar},
  journal={Physica D: nonlinear phenomena},
  volume={9},
  number={1-2},
  pages={189--208},
  year={1983},
  publisher={Elsevier}
}

@article{chang2022evaluating,
  title={Evaluating the different stages of Parkinson’s disease using electroencephalography with holo-Hilbert spectral analysis},
  author={Chang, Kuo-Hsuan and French, Isobel Timothea and Liang, Wei-Kuang and Lo, Yen-Shi and Wang, Yi-Ru and Cheng, Mei-Ling and Huang, Norden E and Wu, Hsiu-Chuan and Lim, Siew-Na and Chen, Chiung-Mei and others},
  journal={Frontiers in aging neuroscience},
  volume={14},
  pages={832637},
  year={2022},
  publisher={Frontiers Media SA}
}

@article{lesmond2005liquidity,
  title={Liquidity of emerging markets},
  author={Lesmond, David A},
  journal={Journal of financial economics},
  volume={77},
  number={2},
  pages={411--452},
  year={2005},
  publisher={Elsevier}
}

@article{hamilton1989new,
  title={A new approach to the economic analysis of nonstationary time series and the business cycle},
  author={Hamilton, James D},
  journal={Econometrica: Journal of the econometric society},
  pages={357--384},
  year={1989},
  publisher={JSTOR}
}

@article{lillo2003power,
  title={Power-law relaxation in a complex system: Omori law after a financial market crash},
  author={Lillo, Fabrizio and Mantegna, Rosario N},
  journal={Physical Review E},
  volume={68},
  number={1},
  pages={016119},
  year={2003},
  publisher={APS}
}

@article{scheffer2009early,
  title={Early-warning signals for critical transitions},
  author={Scheffer, Marten and Bascompte, Jordi and Brock, William A and Brovkin, Victor and Carpenter, Stephen R and Dakos, Vasilis and Held, Hermann and Van Nes, Egbert H and Rietkerk, Max and Sugihara, George},
  journal={Nature},
  volume={461},
  number={7260},
  pages={53--59},
  year={2009},
  publisher={Nature Publishing Group}
}

@article{rai2022statistical,
  title={Statistical properties of the aftershocks of stock market crashes revisited: Analysis based on the 1987 crash, financial-crisis-2008 and COVID-19 pandemic},
  author={Rai, Anish and Mahata, Ajit and Nurujjaman, Md and Prakash, Om},
  journal={International Journal of Modern Physics C},
  volume={33},
  number={02},
  pages={2250019},
  year={2022},
  publisher={World Scientific}
}

@manual{mixvlmc,
  title        = {mixvlmc: Variable Length Markov Chains with Covariates},
  author       = {Rossi, Fabrice and {Le Picard}, Hugo and Joubioux, Gu{\'e}nol{\'e}},
  year         = {2025},
  note         = {{R} package Version 0.2.1.9000},
  organization = {CRAN},
  url          = {https://github.com/fabrice-rossi/mixvlmc},
  urldate      = {2025-12-26}
}

@misc{undesa2025wesp,
  author       = {United Nations Department of Economic and Social Affairs (UN DESA)},
  title        = {World Economic Situation and Prospects 2025},
  year         = {2025},
  url          = {https://desapublications.un.org/publications/world-economic-situation-and-prospects-2025},
  note         = {Accessed: 2025-07-01}
}

@article{lee2022full,
  title={The full informational spectral analysis for auditory steady-state responses in human brain using the combination of canonical correlation analysis and holo-Hilbert spectral analysis},
  author={Lee, Po-Lei and Lee, Te-Min and Lee, Wei-Keung and Chu, Narisa Nan and Shelepin, Yuri E and Hsu, Hao-Teng and Chang, Hsiao-Huang},
  journal={Journal of Clinical Medicine},
  volume={11},
  number={13},
  pages={3868},
  year={2022},
  publisher={MDPI}
}

@article{zheng2023multiscale,
  title={Multiscale three-dimensional Holo--Hilbert spectral entropy: A novel complexity-based early fault feature representation method for rotating machinery},
  author={Zheng, Jinde and Ying, Wanming and Tong, Jinyu and Li, Yongbo},
  journal={Nonlinear Dynamics},
  volume={111},
  number={11},
  pages={10309--10330},
  year={2023},
  publisher={Springer}
}

@article{d2019change,
  title={Change point dynamics for financial data: an indexed Markov chain approach},
  author={D’Amico, Guglielmo and Lika, Ada and Petroni, Filippo},
  journal={Annals of Finance},
  volume={15},
  number={2},
  pages={247--266},
  year={2019},
  publisher={Springer}
}

@article{d2018copula,
  title={Copula based multivariate semi-Markov models with applications in high-frequency finance},
  author={D’Amico, Guglielmo and Petroni, Filippo},
  journal={European Journal of Operational Research},
  volume={267},
  number={2},
  pages={765--777},
  year={2018},
  publisher={Elsevier}
}

@article{d2011semi,
  title={A semi-Markov model with memory for price changes},
  author={D’Amico, Guglielmo and Petroni, Filippo},
  journal={Journal of statistical mechanics: Theory and experiment},
  volume={2011},
  number={12},
  pages={P12009},
  year={2011},
  publisher={IOP Publishing}
}

@article{ying2024order,
  title={Order-frequency holo-Hilbert spectral analysis for machinery fault diagnosis under time-varying operating conditions},
  author={Ying, Wanming and Zheng, Jinde and Huang, Wu and Tong, Jinyu and Pan, Haiyang and Li, Yongbo},
  journal={ISA transactions},
  volume={146},
  pages={472--483},
  year={2024},
  publisher={Elsevier}
}

@article{fantom2016world,
  title={The World Bank's classification of countries by income},
  author={Fantom, Neil James and Serajuddin, Umar},
  journal={World Bank Policy Research Working Paper},
  number={7528},
  year={2016}
}

@article{mcqueen1991stock,
  title={Are stock returns predictable? A test using Markov chains},
  author={McQueen, Grant and Thorley, Steven},
  journal={The Journal of Finance},
  volume={46},
  number={1},
  pages={239--263},
  year={1991},
  publisher={Wiley Online Library}
}

@article{pereda2025systemic,
  title={Systemic Risk and Default Cascades in Global Equity Markets: Extending the Gai-Kapadia Framework with Stochastic Simulations and Network Analysis},
  author={Pereda, Ana IC},
  journal={arXiv preprint arXiv:2504.01969},
  year={2025}
}

@article{li2009tail,
  title={The tail risk of emerging stock markets},
  author={Li, Xiao-Ming and Rose, Lawrence C},
  journal={Emerging markets review},
  volume={10},
  number={4},
  pages={242--256},
  year={2009},
  publisher={Elsevier}
}

@book{embrechts2013modelling,
  title={Modelling extremal events: for insurance and finance},
  author={Embrechts, Paul and Kl{\"u}ppelberg, Claudia and Mikosch, Thomas},
  volume={33},
  year={2013},
  publisher={Springer Science \& Business Media}
}

@article{lamoureux1990persistence,
  title={Persistence in variance, structural change, and the GARCH model},
  author={Lamoureux, Christopher G and Lastrapes, William D},
  journal={Journal of Business \& Economic Statistics},
  volume={8},
  number={2},
  pages={225--234},
  year={1990},
  publisher={Taylor \& Francis}
}

@article{brunnermeier2009deciphering,
  title={Deciphering the liquidity and credit crunch 2007--2008},
  author={Brunnermeier, Markus K},
  journal={Journal of Economic perspectives},
  volume={23},
  number={1},
  pages={77--100},
  year={2009},
  publisher={American Economic Association}
}

@article{shleifer2011fire,
  title={Fire sales in finance and macroeconomics},
  author={Shleifer, Andrei and Vishny, Robert},
  journal={Journal of economic perspectives},
  volume={25},
  number={1},
  pages={29--48},
  year={2011},
  publisher={American Economic Association}
}

@article{engle1982autoregressive,
  title={Autoregressive conditional heteroscedasticity with estimates of the variance of United Kingdom inflation},
  author={Engle, Robert F},
  journal={Econometrica: Journal of the econometric society},
  pages={987--1007},
  year={1982},
  publisher={JSTOR}
}

@article{bollerslev1986generalized,
  title={Generalized autoregressive conditional heteroskedasticity},
  author={Bollerslev, Tim},
  journal={Journal of econometrics},
  volume={31},
  number={3},
  pages={307--327},
  year={1986},
  publisher={Elsevier}
}

@article{cont2001empirical,
  title={Empirical properties of asset returns: stylized facts and statistical issues},
  author={Cont, Rama},
  journal={Quantitative finance},
  volume={1},
  number={2},
  pages={223},
  year={2001},
  publisher={IOP Publishing}
}
\endgroup

\appendix

\section{Appendix}
\label{app:appendix}

\begin{table}[H]
\centering
\caption{BDS test results (embedding dimensions \(m = 2,3\); \(\varepsilon = 0.5\,\sigma\)) for developed and developing market indices. Statistics and p-values are reported separately for each dimension.}
\label{tab:bds}

{\footnotesize
\setlength{\tabcolsep}{6pt}
\renewcommand{\arraystretch}{1.3}
\begin{tabular}{|C{1.7cm}|L{1.7cm}|C{1.4cm}|C{1.4cm}|C{1.4cm}|C{1.4cm}|}
\hline
\multirow{2}{*}{\makecell{Economic\\Condition}}
  & \multirow{2}{*}{\makecell{Market\\Index}}
    & \multicolumn{4}{c|}{BDS test results (at \(m = 2,3\); $\varepsilon = 0.5\sigma$)} \\
\cline{3-6}
  &
    & \makecell{m = 2\\Statistic}
    & \makecell{m = 2\\p-value}
    & \makecell{m = 3\\Statistic}
    & \makecell{m = 3\\p-value} \\
\hline
\multirow{10}{*}{Developed}
  & AXJO      & 2.422 & 0.015 & 3.867 & $<0.001$ \\
  & BFX       & 3.588 & 0.000 & 6.036 & $<0.001$ \\
  & FCHI      & 2.674 & 0.008 & 4.896 & $<0.001$ \\
  & FTSE      & 3.159 & 0.002 & 5.445 & $<0.001$ \\
  & GDAXI     & 2.909 & 0.004 & 5.534 & $<0.001$ \\
  & IBEX      & 2.558 & 0.011 & 4.752 & $<0.001$ \\
  & KS11      & 2.454 & 0.014 & 4.919 & $<0.001$ \\
  & N225      & 1.283 & 0.199 & 2.630 & $<0.01$ \\
  & NYA       & 3.083 & 0.002 & 6.443 & $<0.001$ \\
  & SSMI      & 2.826 & 0.005 & 4.702 & $<0.001$ \\
\hline
\multirow{10}{*}{Developing}
  & BVSP      & 0.814 & 0.416 & 1.753 & 0.08 \\
  & JKSE      & 2.547 & 0.011 & 4.533 & $<0.001$ \\
  & MERV      & 2.160 & 0.031 & 4.027 & $<0.001$ \\
  & MXX       & 2.256 & 0.024 & 3.940 & $<0.001$ \\
  & SET.BK    & 2.548 & 0.011 & 4.865 & $<0.001$ \\
  & STI       & 3.012 & 0.003 & 5.219 & $<0.001$ \\
  & TASI.SR   & 2.994 & 0.003 & 5.866 & $<0.001$ \\
  & TWII      & 1.627 & 0.104 & 3.365 & $<0.001$ \\
  & 000001.SS & 1.882 & 0.060 & 3.705 & $<0.001$ \\
  & 0388.HK   & 2.956 & 0.003 & 4.920 & $<0.001$ \\
\hline
\end{tabular}
}
\end{table}

\begin{figure}[H]
  \centering
  \begin{subfigure}[b]{0.32\linewidth}
    \centering
    \includegraphics[width=\linewidth]{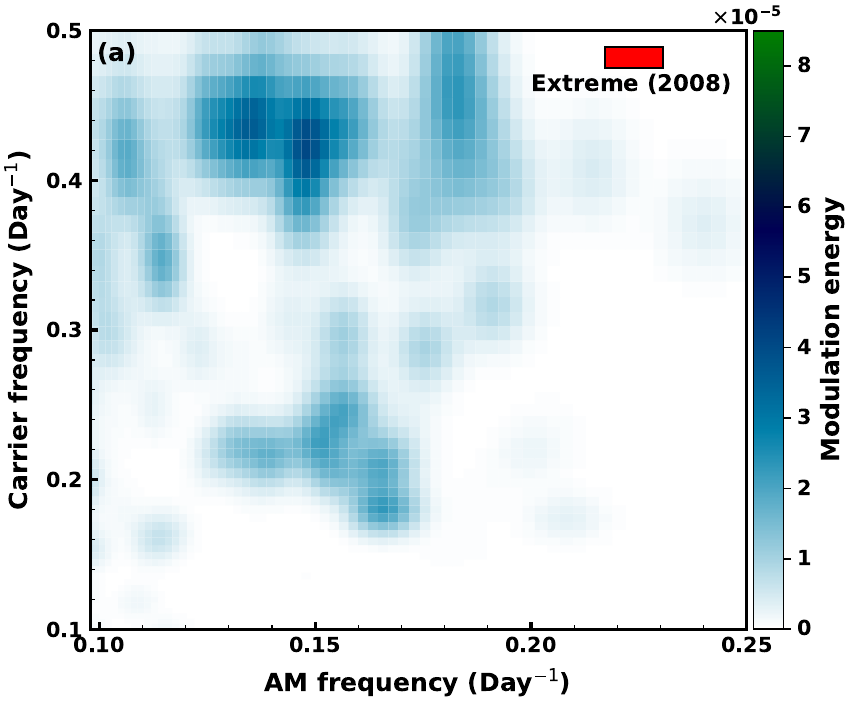}
    \label{fig:2008}
  \end{subfigure}\hfill
  \begin{subfigure}[b]{0.32\linewidth}
    \centering
    \includegraphics[width=\linewidth]{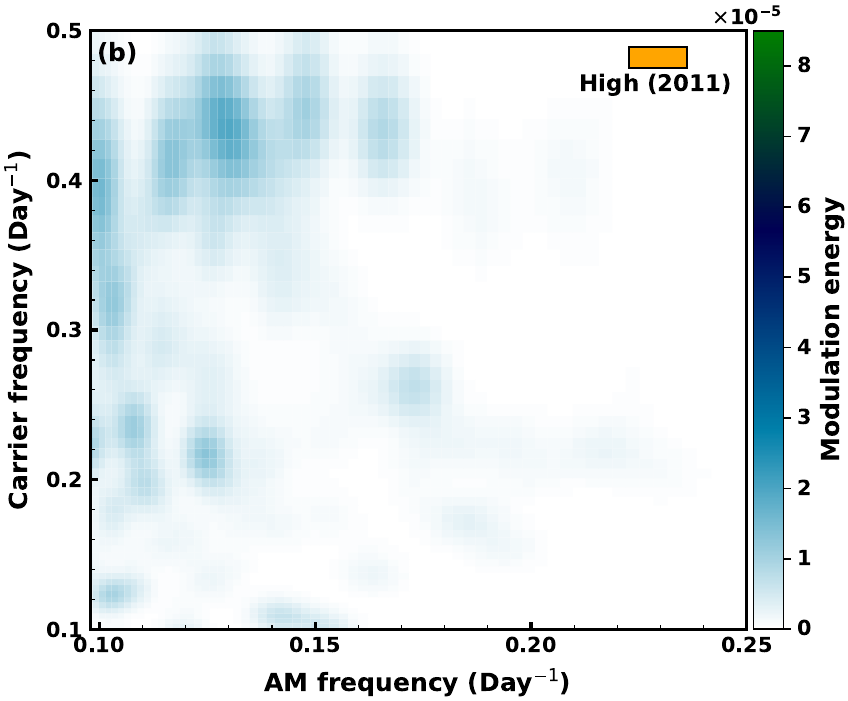}
    \label{fig:2011}
  \end{subfigure}\hfill
  \begin{subfigure}[b]{0.32\linewidth}
    \centering
    \includegraphics[width=\linewidth]{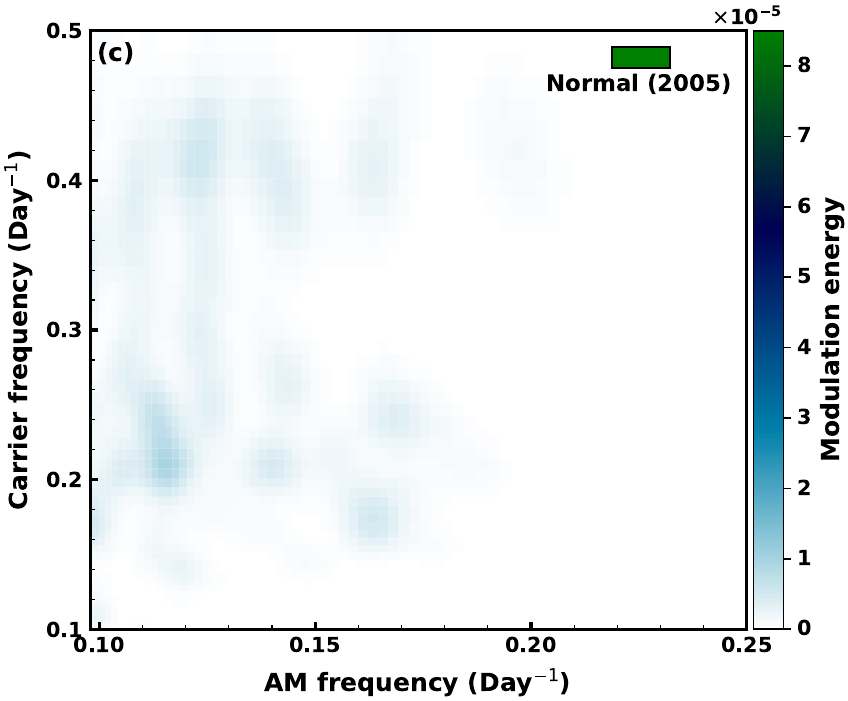}
    \label{fig:2005}
  \end{subfigure}
  \caption{Holo--Hilbert spectra (HHS) for the Bovespa index (BVSP) over one-year windows selected to represent the three regimes identified from the instantaneous energy series: (a) Extreme regime year 2008, corresponding to the red-coded energy points, (b) High regime year 2011, corresponding to the orange-coded energy points, and (c) Normal regime year 2005, corresponding to the green-coded energy points. In each panel, the vertical axis is the carrier frequency $\omega_c$ and the horizontal axis is the amplitude-modulation frequency $\omega_{am}$, while the color scale indicates amplitude-modulation energy, used here as a measure of volatility intensity.}
  \label{fig:HHS_panels_BVSP}
\end{figure}

\begin{table}[H]
\centering
\footnotesize
\setlength{\tabcolsep}{3pt}
\begin{tabular}{|l|p{3.5cm}|p{5.3cm}|p{5.3cm}|}
\hline
Index & \multicolumn{3}{c|}{Regime} \\
\cline{2-4}
 & Extreme & High & Normal \\
\hline
AXJO & 2008, 2020 & 2000, 2001, 2007, 2009, 2010, 2011, 2013, 2015, 2016, 2022, 2025 & 2002, 2003, 2004, 2005, 2006, 2012, 2014, 2017, 2018, 2019, 2021, 2023, 2024 \\
\hline
BFX & 2003, 2008, 2010, 2020 & 2000, 2001, 2002, 2007, 2009, 2011, 2015, 2016, 2022, 2025 & 2004, 2005, 2006, 2012, 2013, 2014, 2017, 2018, 2019, 2021, 2023, 2024 \\
\hline
FCHI & 2002, 2008, 2009, 2010, 2020 & 2000, 2001, 2003, 2011, 2012, 2015, 2016, 2021, 2022, 2025 & 2004, 2005, 2006, 2007, 2013, 2014, 2017, 2018, 2019, 2023, 2024 \\
\hline
FTSE & 2008, 2020 & 2000, 2001, 2002, 2003, 2007, 2009, 2010, 2011, 2015, 2016, 2022, 2025 & 2004, 2005, 2006, 2012, 2013, 2014, 2017, 2018, 2019, 2021, 2023, 2024 \\
\hline
GDAXI & 2002, 2008, 2020 & 2000, 2001, 2003, 2009, 2010, 2011, 2012, 2013, 2015, 2016, 2022, 2025 & 2004, 2005, 2006, 2007, 2014, 2017, 2018, 2019, 2021, 2023, 2024 \\
\hline
IBEX & 2008, 2010, 2016, 2020 & 2000, 2001, 2002, 2003, 2009, 2011, 2012, 2015, 2022, 2023, 2025 & 2004, 2005, 2006, 2007, 2013, 2014, 2017, 2018, 2019, 2021, 2024 \\
\hline
KS11 & 2000, 2001, 2008, 2020 & 2002, 2003, 2004, 2007, 2009, 2011, 2021, 2023, 2024, 2025 & 2005, 2006, 2010, 2012, 2013, 2014, 2015, 2016, 2017, 2018, 2019, 2022 \\
\hline
N225 & 2008, 2011, 2024, 2025 & 2000, 2001, 2002, 2003, 2007, 2009, 2010, 2013, 2014, 2015, 2016, 2018, 2020, 2022 & 2004, 2005, 2006, 2012, 2017, 2019, 2021, 2023 \\
\hline
NYA & 2008, 2020 & 2000, 2001, 2002, 2007, 2009, 2010, 2011, 2015, 2016, 2018, 2022, 2025 & 2003, 2004, 2005, 2006, 2012, 2013, 2014, 2017, 2019, 2021, 2023, 2024 \\
\hline
SSMI & 2001, 2008, 2020 & 2000, 2002, 2003, 2007, 2009, 2010, 2011, 2015, 2016, 2022, 2025 & 2004, 2005, 2006, 2012, 2013, 2014, 2017, 2018, 2019, 2021, 2023, 2024 \\
\hline
\end{tabular}
\caption{Regime years for developed stock market indices.}
\label{tab:regime_years_developed}
\end{table}

\begin{table}[H]
\centering
\footnotesize
\setlength{\tabcolsep}{3pt}
\begin{tabular}{|l|p{3.5cm}|p{5.3cm}|p{5.3cm}|}
\hline
Index & \multicolumn{3}{c|}{Regime} \\
\cline{2-4}
 & Extreme & High & Normal \\
\hline
000001.SS & 2001, 2007, 2008, 2015 & 2000, 2002, 2003, 2004, 2005, 2006, 2009, 2010, 2012, 2013, 2014, 2016, 2018, 2019, 2020, 2022, 2024, 2025 & 2011, 2017, 2021, 2023 \\
\hline
0388.HK & 2001, 2007, 2008, 2015, 2024 & 2000, 2002, 2003, 2004, 2006, 2009, 2010, 2011, 2012, 2014, 2016, 2018, 2020, 2021, 2022, 2025 & 2005, 2013, 2017, 2019, 2023 \\
\hline
BVSP & 2008, 2020 & 2000, 2001, 2002, 2004, 2006, 2007, 2009, 2011, 2014, 2016, 2017, 2021 & 2003, 2005, 2010, 2012, 2013, 2015, 2018, 2019, 2022, 2023, 2024, 2025 \\
\hline
JKSE & 2002, 2004, 2006, 2007, 2008, 2011, 2020, 2025 & 2000, 2001, 2003, 2005, 2009, 2010, 2012, 2013, 2015, 2018, 2022 & 2014, 2016, 2017, 2019, 2021, 2023, 2024 \\
\hline
MERV & 2019 & 2001, 2002, 2004, 2008, 2011, 2014, 2018, 2020, 2023, 2024, 2025 & 2000, 2003, 2005, 2006, 2007, 2009, 2010, 2012, 2013, 2015, 2016, 2017, 2021, 2022 \\
\hline
MXX & 2000, 2008 & 2001, 2002, 2004, 2006, 2007, 2009, 2011, 2016, 2018, 2020, 2024, 2025 & 2003, 2005, 2010, 2012, 2013, 2014, 2015, 2017, 2019, 2021, 2022, 2023 \\
\hline
SET.BK & 2006, 2008, 2020 & 2000, 2001, 2002, 2004, 2007, 2009, 2010, 2011, 2013, 2015, 2016, 2025 & 2003, 2005, 2012, 2014, 2017, 2018, 2019, 2021, 2022, 2023, 2024 \\
\hline
STI & 2000, 2008, 2020 & 2001, 2002, 2003, 2006, 2007, 2009, 2010, 2011, 2015, 2016, 2018, 2024, 2025 & 2004, 2005, 2012, 2013, 2014, 2017, 2019, 2021, 2022, 2023 \\
\hline
TASI.SR & 2006, 2008, 2015 & 2003, 2004, 2005, 2007, 2009, 2010, 2011, 2014, 2018, 2020, 2025 & 2000, 2001, 2002, 2012, 2013, 2016, 2017, 2019, 2021, 2022, 2023, 2024 \\
\hline
TWII & 2000, 2008, 2009, 2020, 2024, 2025 & 2001, 2002, 2003, 2004, 2006, 2007, 2010, 2011, 2015, 2016, 2018, 2021, 2022 & 2005, 2012, 2013, 2014, 2017, 2019, 2023 \\
\hline
\end{tabular}
\caption{Regime years for developing stock market indices.}
\label{tab:regime_years_developing}
\end{table}

\subsection*{Sensitivity analysis of regime thresholds}
\label{appendix:sensitivity}

Regime identification in this study is based on thresholding the normalized instantaneous energy series \(E(t)\) obtained from the Hilbert--Huang Transform. The baseline specification classifies observations using two cutoffs,
\[
\tau_1=\mu+\sigma,
\qquad
\tau_2=\mu+6\sigma,
\]
where \(\mu\) and \(\sigma\) denote the sample mean and standard deviation of \(E(t)\), respectively. Observations are assigned to Normal, High, and Extreme regimes according to
\[
\text{Normal: } E(t)\le\tau_1,\qquad
\text{High: } \tau_1 < E(t)\le\tau_2,\qquad
\text{Extreme: } E(t)>\tau_2.
\]
To assess robustness to threshold choice, we perform a grid-based sensitivity analysis by perturbing both cutoffs via multiplicative factors \(a\) and \(b\):
\[
\tau_1(a)=\mu+a\sigma,
\qquad
\tau_2(b)=\mu+b\sigma,
\]
and re-running the regime assignment for each \((a,b)\) pair in the grid
\[
a\in\{0.75,\,1.00,\,1.25\},
\qquad
b\in\{4.5,\,6.0,\,7.5\},
\qquad b>a.
\]
This design retains the baseline case \((a,b)=(1,6)\) while spanning both a less conservative and a more conservative separation of regimes. For each index and each \((a,b)\) pair, daily observations are first classified into regimes by \(\tau_1(a)\) and \(\tau_2(b)\). The corresponding regime years are then defined as the set of calendar years that contain at least one day assigned to that regime. To ensure that each year belongs to only one regime for a given index, overlaps are removed by a severity rule: if a year is flagged as Extreme, it is excluded from High and Normal; if a year is flagged as High, it is excluded from Normal.

Sensitivity is evaluated separately for developed and developing market panels. For each \((a,b)\) configuration and regime, we compute the intersection of regime-year sets across indices within a panel. When the intersection is empty for a regime, we report a fallback set consisting of the two most frequently occurring regime years across indices in that panel. Robustness is then assessed by comparing the regime-year sets obtained under each \((a,b)\) configuration to those under the baseline thresholds. We summarize similarity using set-overlap measures such as the Jaccard similarity, and by verifying that the qualitative ordering of regimes is preserved, namely that Extreme years remain concentrated in crisis episodes, High years correspond to sustained elevated volatility periods, and Normal years correspond to comparatively tranquil conditions. Detailed results of the sensitivity analysis are available on Figshare at \href{https://doi.org/10.6084/m9.figshare.30982552}{10.6084/m9.figshare.30982552}. Consistency of these conclusions across the threshold grid supports the stability of the regime identification procedure.

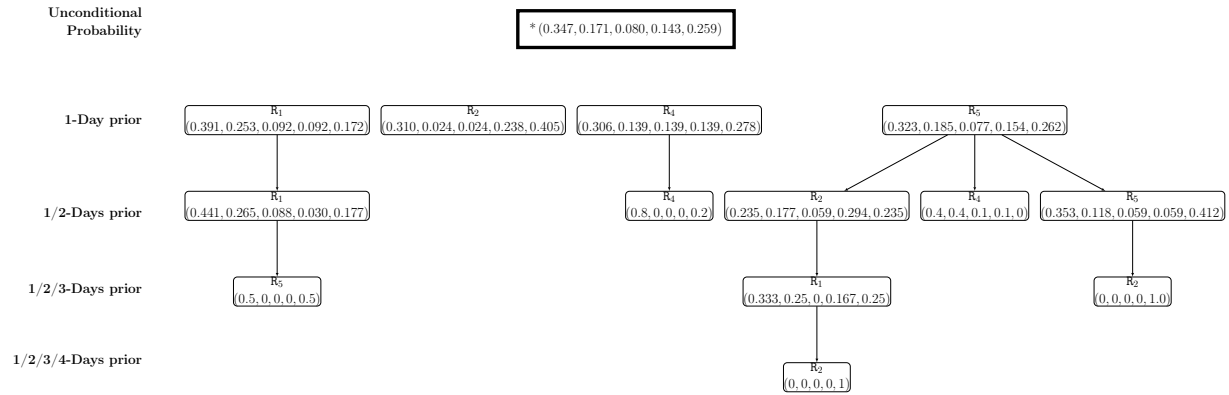
\begin{figure}[H]
    \centering
    \resizebox{\textwidth}{!}{%
    \begin{tikzpicture}[
        daylabel/.style={font=\large, anchor=east, align=right}, 
        scale=0.99,
        transform shape
    ]
        \node (tree) [inner sep=0pt]  {
            \begin{forest}
                for tree={
                    draw,
                    rounded corners,
                    node options={align=center,font=\large},
                    edge={-latex},
                    s sep=1em,
                    l sep=5.0em,
                    anchor=north,
                    grow=south,
                    where level=1{no edge}{}
                },
                for root={
                    rectangle,
                    sharp corners,
                    line width=3.5pt,
                    inner sep=12pt
                }
                [
                  {*\,(0.347,\,0.171,\,0.080,\,0.143,\,0.259)}
                    [
                      {$\mathtt{R}_1$\\(0.391,\,0.253,\,0.092,\,0.092,\,0.172)}
                      [
                        {$\mathtt{R}_1$\\(0.441,\,0.265,\,0.088,\,0.030,\,0.177)}
                        [
                          {$\mathtt{R}_5$\\(0.5,\,0,\,0,\,0,\,0.5)}
                        ]
                      ]
                    ]
                    [
                      {$\mathtt{R}_2$\\(0.310,\,0.024,\,0.024,\,0.238,\,0.405)}
                    ]
                    [
                      {$\mathtt{R}_4$\\(0.306,\,0.139,\,0.139,\,0.139,\,0.278)}
                      [
                        {$\mathtt{R}_4$\\(0.8,\,0,\,0,\,0,\,0.2)}
                      ]
                    ]
                    [
                      {$\mathtt{R}_5$\\(0.323,\,0.185,\,0.077,\,0.154,\,0.262)}
                      [
                        {$\mathtt{R}_2$\\(0.235,\,0.177,\,0.059,\,0.294,\,0.235)}
                        [
                          {$\mathtt{R}_1$\\(0.333,\,0.25,\,0,\,0.167,\,0.25)}
                          [
                            {$\mathtt{R}_2$\\(0,\,0,\,0,\,0,\,1)}
                          ]
                        ]
                      ]
                      [
                        {$\mathtt{R}_4$\\(0.4,\,0.4,\,0.1,\,0.1,\,0)}
                      ]
                      [
                        {$\mathtt{R}_5$\\(0.353,\,0.118,\,0.059,\,0.059,\,0.412)}
                        [
                          {$\mathtt{R}_2$\\(0,\,0,\,0,\,0,\,1.0)}
                        ]
                      ]
                    ]
                ]
            \end{forest}
        };

        \path (tree.north west) ++(-1.4cm,-0.5cm)  node[daylabel] {\textbf{Unconditional}\\\textbf{Probability}};
        \path (tree.north west) ++(-1.4cm,-4.1cm)  node[daylabel] {\textbf{1-Day prior}};
        \path (tree.north west) ++(-1.4cm,-7.5cm)  node[daylabel] {\textbf{1/2-Days prior}};
        \path (tree.north west) ++(-1.4cm,-10.3cm)  node[daylabel] {\textbf{1/2/3-Days prior}};
        \path (tree.north west) ++(-1.4cm,-12.9cm)  node[daylabel] {\textbf{1/2/3/4-Days prior}};
    \end{tikzpicture}%
    }
    \caption{High (2022) regime context tree for NYSE Composite (NYA) index. The root node (*) a the bold square border shows unconditional probabilities of $\mathtt{R}_1, \mathtt{R}_2, \mathtt{R}_3, \mathtt{R}_4 \hspace{2pt} \text{and}\hspace{2pt} \mathtt{R}_5$.}
    \label{fig:vlmc-case2}
\end{figure}

\begin{figure}[H]
    \centering
    \resizebox{0.9\textwidth}{!}{%
    \begin{tikzpicture}[
        daylabel/.style={font=\small, anchor=east, align=right}, 
        scale=0.95,
        transform shape
    ]
        \node (tree) [inner sep=0pt]  {
            \begin{forest}
                for tree={
                    draw,
                    rounded corners,
                    node options={align=center,font=\footnotesize},
                    edge={-latex},
                    s sep=1em,
                    l sep=3.0em,
                    anchor=north,
                    grow=south,
                    where level=1{no edge}{}
                },
                for root={
                    rectangle,
                    sharp corners,
                    line width=2.5pt,
                    inner sep=6pt
                }
                [
                  {*\,(0.183,\,0.234,\,0.202,\,0.258,\,0.123)}
                    [
                      {$\mathtt{R}_1$\\(0.217,\,0.174,\,0.217,\,0.217,\,0.174)}
                      [
                        {$\mathtt{R}_3$\\(0.143,\,0.143,\,0.143,\,0.286,\,0.286)}
                        [
                          {$\mathtt{R}_2$\\(0,\,0.4,\,0,\,0.6,\,0)}
                        ]
                      ]
                    ]
                    [
                      {$\mathtt{R}_2$\\(0.138,\,0.138,\,0.345,\,0.224,\,0.155)}
                      [
                        {$\mathtt{R}_5$\\(0.3,\,0.3,\,0.2,\,0,\,0.2)}
                      ]
                    ]
                    [
                      {$\mathtt{R}_3$\\(0.275,\,0.255,\,0.157,\,0.196,\,0.118)}
                      [
                        {$\mathtt{R}_4$\\(0.5,\,0.5,\,0,\,0,\,0)}
                      ]
                      [
                        {$\mathtt{R}_5$\\(0.143,\,0.714,\,0.143,\,0,\,0)}
                      ]
                    ]
                    [
                      {$\mathtt{R}_4$\\(0.139,\,0.308,\,0.092,\,0.369,\,0.092)}
                    ]
                ]
            \end{forest}
        };

        \path (tree.north west) ++(-1.4cm,-0.4cm)  node[daylabel] {\textbf{Unconditional}\\\textbf{Probability}};
        \path (tree.north west) ++(-1.4cm,-2.6cm)  node[daylabel] {\textbf{1-Day prior}};
        \path (tree.north west) ++(-1.4cm,-4.5cm)  node[daylabel] {\textbf{1/2-Days prior}};
        \path (tree.north west) ++(-1.4cm,-6.4cm)  node[daylabel] {\textbf{1/2/3-Days prior}};
    \end{tikzpicture}%
    }
    \caption{Normal (2005) regime context tree for NYSE Composite (NYA) index. The root node (*) a the bold square border shows unconditional probabilities of $\mathtt{R}_1, \mathtt{R}_2, \mathtt{R}_3, \mathtt{R}_4 \hspace{2pt} \text{and}\hspace{2pt} \mathtt{R}_5$.}
    \label{fig:vlmc-case3}
\end{figure}
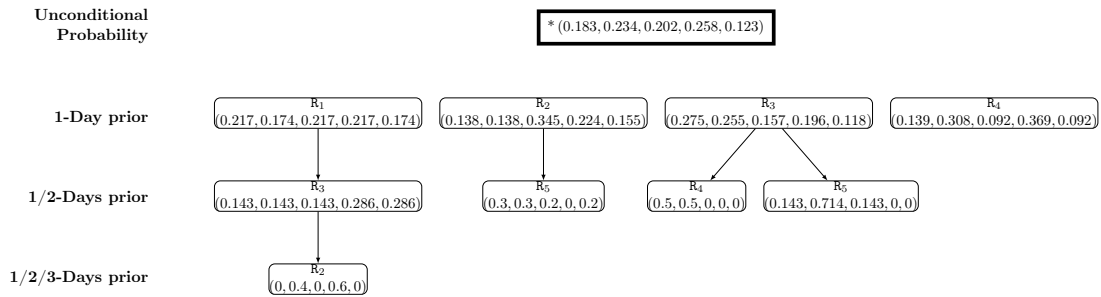


\begin{table}[H]
\centering
\scriptsize
\setlength{\tabcolsep}{3pt}
\renewcommand{\arraystretch}{1.2}

\begin{tabular}{cc}

\begin{minipage}[t]{0.48\linewidth}
\centering
\caption{Contexts with Count $>2$ for developed stock market indices during High regimes.}
\label{tab:developed-high}
\begin{tabular}{|l|c|*{5}{>{\centering\arraybackslash}p{0.85cm}|}}
\hline
\multirow{2}{*}{Context} & \multirow{2}{*}{Count} & \multicolumn{5}{c|}{Probabilities to after (State)} \\
\cline{3-7}
 & & P$(\mathtt{R}_1)$ & P$(\mathtt{R}_2)$ & P$(\mathtt{R}_3)$ & P$(\mathtt{R}_4)$ & P$(\mathtt{R}_5)$ \\
\hline
$\mathtt{R}_1$ & 13 & 0.277 & 0.196 & 0.137 & 0.163 & 0.227 \\
$\mathtt{R}_1\mathtt{R}_1$ & 5 & 0.282 & 0.169 & 0.151 & 0.182 & 0.216 \\
$\mathtt{R}_1\mathtt{R}_1\mathtt{R}_5$ & 3 & 0.250 & 0.163 & 0.150 & 0.188 & 0.250 \\
$\mathtt{R}_1\mathtt{R}_3$ & 4 & 0.312 & 0.327 & 0.102 & 0.161 & 0.099 \\
$\mathtt{R}_1\mathtt{R}_4$ & 6 & 0.231 & 0.362 & 0.085 & 0.119 & 0.203 \\
\hline
$\mathtt{R}_2$ & 12 & 0.262 & 0.170 & 0.143 & 0.206 & 0.220 \\
$\mathtt{R}_2\mathtt{R}_1$ & 3 & 0.078 & 0.171 & 0.279 & 0.075 & 0.398 \\
$\mathtt{R}_2\mathtt{R}_4$ & 8 & 0.254 & 0.188 & 0.226 & 0.264 & 0.068 \\
$\mathtt{R}_2\mathtt{R}_5$ & 5 & 0.156 & 0.296 & 0.183 & 0.109 & 0.256 \\
\hline
$\mathtt{R}_3$ & 9 & 0.230 & 0.218 & 0.177 & 0.210 & 0.166 \\
$\mathtt{R}_3\mathtt{R}_3$ & 3 & 0.204 & 0.256 & 0.098 & 0.238 & 0.204 \\
\hline
$\mathtt{R}_4$ & 17 & 0.224 & 0.221 & 0.189 & 0.185 & 0.181 \\
$\mathtt{R}_4\mathtt{R}_1$ & 6 & 0.266 & 0.072 & 0.109 & 0.191 & 0.362 \\
$\mathtt{R}_4\mathtt{R}_2$ & 7 & 0.312 & 0.090 & 0.177 & 0.246 & 0.175 \\
$\mathtt{R}_4\mathtt{R}_3$ & 6 & 0.266 & 0.210 & 0.190 & 0.169 & 0.165 \\
$\mathtt{R}_4\mathtt{R}_4$ & 5 & 0.283 & 0.252 & 0.208 & 0.137 & 0.119 \\
\hline
$\mathtt{R}_5$ & 12 & 0.238 & 0.197 & 0.182 & 0.182 & 0.202 \\
$\mathtt{R}_5\mathtt{R}_1$ & 4 & 0.253 & 0.167 & 0.155 & 0.209 & 0.218 \\
$\mathtt{R}_5\mathtt{R}_2$ & 5 & 0.191 & 0.219 & 0.102 & 0.281 & 0.207 \\
$\mathtt{R}_5\mathtt{R}_3$ & 3 & 0.268 & 0.292 & 0.250 & 0.036 & 0.155 \\
$\mathtt{R}_5\mathtt{R}_4$ & 4 & 0.102 & 0.323 & 0.142 & 0.320 & 0.113 \\
$\mathtt{R}_5\mathtt{R}_5$ & 3 & 0.284 & 0.123 & 0.145 & 0.270 & 0.179 \\
\hline
\end{tabular}
\end{minipage}

& 

\begin{minipage}[t]{0.48\linewidth}
\centering
\caption{Contexts with Count $>2$ for developed stock market indices during Normal regimes.}
\label{tab:developed-normal}
\begin{tabular}{|l|c|*{5}{>{\centering\arraybackslash}p{0.85cm}|}}
\hline
\multirow{2}{*}{Context} & \multirow{2}{*}{Count} & \multicolumn{5}{c|}{Probability to after (State)} \\
\cline{3-7}
 & & P$(\mathtt{R}_1)$ & P$(\mathtt{R}_2)$ & P$(\mathtt{R}_3)$ & P$(\mathtt{R}_4)$ & P$(\mathtt{R}_5)$ \\
\hline
$\mathtt{R}_1$ & 8 & 0.182 & 0.244 & 0.194 & 0.214 & 0.166 \\
$\mathtt{R}_1\mathtt{R}_1$ & 3 & 0.083 & 0.188 & 0.000 & 0.188 & 0.542 \\
$\mathtt{R}_1\mathtt{R}_2$ & 4 & 0.098 & 0.320 & 0.120 & 0.365 & 0.097 \\
\hline
$\mathtt{R}_2$ & 15 & 0.117 & 0.260 & 0.260 & 0.257 & 0.106 \\
$\mathtt{R}_2\mathtt{R}_1$ & 3 & 0.108 & 0.000 & 0.125 & 0.558 & 0.208 \\
$\mathtt{R}_2\mathtt{R}_2$ & 10 & 0.138 & 0.249 & 0.229 & 0.269 & 0.116 \\
$\mathtt{R}_2\mathtt{R}_3$ & 3 & 0.119 & 0.192 & 0.335 & 0.204 & 0.151 \\
$\mathtt{R}_2\mathtt{R}_4$ & 6 & 0.148 & 0.258 & 0.230 & 0.256 & 0.109 \\
\hline
$\mathtt{R}_3$ & 15 & 0.105 & 0.295 & 0.270 & 0.232 & 0.098 \\
$\mathtt{R}_3\mathtt{R}_2$ & 3 & 0.069 & 0.228 & 0.229 & 0.330 & 0.144 \\
$\mathtt{R}_3\mathtt{R}_3$ & 4 & 0.200 & 0.172 & 0.201 & 0.254 & 0.173 \\
$\mathtt{R}_3\mathtt{R}_4$ & 9 & 0.156 & 0.367 & 0.252 & 0.178 & 0.048 \\
$\mathtt{R}_3\mathtt{R}_5$ & 5 & 0.027 & 0.246 & 0.280 & 0.358 & 0.089 \\
\hline
$\mathtt{R}_4$ & 15 & 0.095 & 0.309 & 0.295 & 0.209 & 0.093 \\
$\mathtt{R}_4\mathtt{R}_2$ & 5 & 0.176 & 0.308 & 0.189 & 0.225 & 0.104 \\
$\mathtt{R}_4\mathtt{R}_2\mathtt{R}_3$ & 3 & 0.200 & 0.333 & 0.000 & 0.233 & 0.233 \\
$\mathtt{R}_4\mathtt{R}_4$ & 4 & 0.137 & 0.381 & 0.168 & 0.237 & 0.077 \\
$\mathtt{R}_4\mathtt{R}_5$ & 6 & 0.048 & 0.487 & 0.376 & 0.078 & 0.012 \\
\hline
$\mathtt{R}_5$ & 7 & 0.046 & 0.247 & 0.316 & 0.286 & 0.105 \\
\hline
\end{tabular}
\end{minipage}

\end{tabular}
\end{table}

\begin{table}[H]
\centering
\scriptsize                    
\setlength{\tabcolsep}{4pt}      
\renewcommand{\arraystretch}{1.1}

\begin{tabular}{cc}             

\begin{minipage}[t]{0.48\linewidth}
\centering
\caption{Contexts with Count $>2$ for developing stock market indices during Extreme regimes.}
\label{tab:developing-extreme}
\begin{tabular}{|l|c|r|r|r|r|r|}
\hline
\multirow{2}{*}{Context} & \multirow{2}{*}{Count} & \multicolumn{5}{c|}{Probabilities to after (State)} \\
\cline{3-7}
 &  & P($\mathtt{R}_1$) & P($\mathtt{R}_2$) & P($\mathtt{R}_3$) & P($\mathtt{R}_4$) & P($\mathtt{R}_5$) \\
\hline
$\mathtt{R}_1$ & 19 & 0.336 & 0.139 & 0.131 & 0.137 & 0.258 \\
$\mathtt{R}_1\mathtt{R}_1$ & 6 & 0.399 & 0.094 & 0.074 & 0.139 & 0.295 \\
$\mathtt{R}_1\mathtt{R}_2$ & 7 & 0.228 & 0.117 & 0.283 & 0.142 & 0.231 \\
$\mathtt{R}_1\mathtt{R}_3$ & 6 & 0.386 & 0.272 & 0.128 & 0.072 & 0.142 \\
$\mathtt{R}_1\mathtt{R}_4$ & 9 & 0.396 & 0.207 & 0.133 & 0.136 & 0.128 \\
$\mathtt{R}_1\mathtt{R}_5$ & 6 & 0.343 & 0.123 & 0.127 & 0.135 & 0.272 \\
\hline
$\mathtt{R}_2$ & 13 & 0.286 & 0.188 & 0.161 & 0.154 & 0.211 \\
$\mathtt{R}_2\mathtt{R}_1$ & 3 & 0.304 & 0.149 & 0.201 & 0.126 & 0.220 \\
$\mathtt{R}_2\mathtt{R}_2$ & 3 & 0.188 & 0.283 & 0.351 & 0.000 & 0.178 \\
$\mathtt{R}_2\mathtt{R}_4$ & 3 & 0.504 & 0.159 & 0.115 & 0.124 & 0.099 \\
\hline
$\mathtt{R}_3$ & 9 & 0.239 & 0.192 & 0.168 & 0.191 & 0.210 \\
$\mathtt{R}_3\mathtt{R}_3$ & 4 & 0.153 & 0.104 & 0.086 & 0.124 & 0.533 \\
$\mathtt{R}_3\mathtt{R}_5$ & 3 & 0.022 & 0.262 & 0.484 & 0.118 & 0.115 \\
\hline
$\mathtt{R}_4$ & 13 & 0.272 & 0.211 & 0.133 & 0.171 & 0.213 \\
$\mathtt{R}_4\mathtt{R}_3$ & 3 & 0.283 & 0.257 & 0.179 & 0.260 & 0.023 \\
$\mathtt{R}_4\mathtt{R}_4$ & 4 & 0.000 & 0.077 & 0.375 & 0.307 & 0.241 \\
$\mathtt{R}_4\mathtt{R}_5$ & 3 & 0.321 & 0.177 & 0.033 & 0.067 & 0.403 \\
\hline
$\mathtt{R}_5$ & 18 & 0.276 & 0.174 & 0.163 & 0.134 & 0.253 \\
$\mathtt{R}_5\mathtt{R}_1$ & 8 & 0.326 & 0.199 & 0.129 & 0.095 & 0.252 \\
$\mathtt{R}_5\mathtt{R}_1\mathtt{R}_1$ & 4 & 0.431 & 0.000 & 0.049 & 0.361 & 0.160 \\
$\mathtt{R}_5\mathtt{R}_3$ & 4 & 0.175 & 0.270 & 0.160 & 0.036 & 0.360 \\
$\mathtt{R}_5\mathtt{R}_5$ & 6 & 0.255 & 0.181 & 0.127 & 0.144 & 0.294 \\
\hline
\end{tabular}
\end{minipage}
&
\begin{minipage}[t]{0.48\linewidth}
\centering
\caption{Contexts with Count $>2$ for developing stock market indices during High regimes.}
\label{tab:developing-high}
\begin{tabular}{|l|c|r|r|r|r|r|}
\hline
\multirow{2}{*}{Context} & \multirow{2}{*}{Count} & \multicolumn{5}{c|}{Probabilities to after (State)} \\
\cline{3-7}
 &  & P($\mathtt{R}_1$) & P($\mathtt{R}_2$) & P($\mathtt{R}_3$) & P($\mathtt{R}_4$) & P($\mathtt{R}_5$) \\
\hline
$\mathtt{R}_1$ & 18 & 0.295 & 0.166 & 0.161 & 0.144 & 0.235 \\
$\mathtt{R}_1\mathtt{R}_1$ & 4 & 0.195 & 0.357 & 0.060 & 0.046 & 0.342 \\
$\mathtt{R}_1\mathtt{R}_2$ & 4 & 0.336 & 0.213 & 0.208 & 0.096 & 0.147 \\
$\mathtt{R}_1\mathtt{R}_3$ & 6 & 0.401 & 0.071 & 0.239 & 0.167 & 0.122 \\
$\mathtt{R}_1\mathtt{R}_4$ & 5 & 0.145 & 0.238 & 0.182 & 0.263 & 0.172 \\
$\mathtt{R}_1\mathtt{R}_5$ & 4 & 0.238 & 0.195 & 0.330 & 0.080 & 0.157 \\
\hline
$\mathtt{R}_2$ & 13 & 0.278 & 0.185 & 0.184 & 0.139 & 0.214 \\
$\mathtt{R}_2\mathtt{R}_1$ & 4 & 0.239 & 0.177 & 0.174 & 0.148 & 0.263 \\
$\mathtt{R}_2\mathtt{R}_3$ & 4 & 0.091 & 0.210 & 0.301 & 0.162 & 0.237 \\
$\mathtt{R}_2\mathtt{R}_5$ & 5 & 0.208 & 0.193 & 0.243 & 0.149 & 0.207 \\
\hline
$\mathtt{R}_3$ & 8 & 0.293 & 0.143 & 0.156 & 0.170 & 0.237 \\
$\mathtt{R}_3\mathtt{R}_2$ & 4 & 0.208 & 0.055 & 0.345 & 0.144 & 0.249 \\
\hline
$\mathtt{R}_4$ & 13 & 0.225 & 0.212 & 0.170 & 0.178 & 0.214 \\
$\mathtt{R}_4\mathtt{R}_1$ & 3 & 0.174 & 0.000 & 0.278 & 0.261 & 0.288 \\
$\mathtt{R}_4\mathtt{R}_2$ & 6 & 0.081 & 0.254 & 0.175 & 0.412 & 0.078 \\
$\mathtt{R}_4\mathtt{R}_3$ & 3 & 0.127 & 0.360 & 0.267 & 0.083 & 0.163 \\
$\mathtt{R}_4\mathtt{R}_5$ & 5 & 0.287 & 0.194 & 0.245 & 0.130 & 0.144 \\
\hline
$\mathtt{R}_5$ & 15 & 0.225 & 0.187 & 0.170 & 0.194 & 0.225 \\
$\mathtt{R}_5\mathtt{R}_1$ & 3 & 0.303 & 0.303 & 0.101 & 0.106 & 0.187 \\
$\mathtt{R}_5\mathtt{R}_2$ & 4 & 0.214 & 0.113 & 0.218 & 0.155 & 0.301 \\
$\mathtt{R}_5\mathtt{R}_3$ & 7 & 0.165 & 0.165 & 0.329 & 0.240 & 0.101 \\
$\mathtt{R}_5\mathtt{R}_4$ & 5 & 0.200 & 0.231 & 0.155 & 0.307 & 0.108 \\
$\mathtt{R}_5\mathtt{R}_5$ & 5 & 0.228 & 0.122 & 0.179 & 0.159 & 0.313 \\
$\mathtt{R}_5\mathtt{R}_5\mathtt{R}_5$ & 3 & 0.393 & 0.322 & 0.000 & 0.286 & 0.000 \\
\hline
\end{tabular}
\end{minipage}
\\[13.5cm]

\multicolumn{2}{c}{
\begin{minipage}[t]{0.70\linewidth}
\centering
\caption{Contexts with Count $>2$ for developing stock market indices during Normal regimes.}
\label{tab:developing-normal}
\begin{tabular}{|l|c|r|r|r|r|r|}
\hline
\multirow{2}{*}{Context} & \multirow{2}{*}{Count} & \multicolumn{5}{c|}{Probabilities to after (State)} \\
\cline{3-7}
 &  & P($\mathtt{R}_1$) & P($\mathtt{R}_2$) & P($\mathtt{R}_3$) & P($\mathtt{R}_4$) & P($\mathtt{R}_5$) \\
\hline
$\mathtt{R}_1$ & 8 & 0.254 & 0.161 & 0.190 & 0.248 & 0.147 \\
\hline
$\mathtt{R}_2$ & 14 & 0.142 & 0.242 & 0.282 & 0.238 & 0.096 \\
$\mathtt{R}_2\mathtt{R}_2$ & 6 & 0.072 & 0.328 & 0.215 & 0.221 & 0.165 \\
$\mathtt{R}_2\mathtt{R}_3$ & 5 & 0.119 & 0.265 & 0.255 & 0.260 & 0.102 \\
$\mathtt{R}_2\mathtt{R}_4$ & 3 & 0.094 & 0.266 & 0.255 & 0.229 & 0.156 \\
$\mathtt{R}_2\mathtt{R}_5$ & 3 & 0.176 & 0.307 & 0.327 & 0.042 & 0.149 \\
\hline
$\mathtt{R}_3$ & 14 & 0.140 & 0.260 & 0.281 & 0.201 & 0.118 \\
$\mathtt{R}_3\mathtt{R}_1$ & 6 & 0.130 & 0.247 & 0.270 & 0.188 & 0.165 \\
$\mathtt{R}_3\mathtt{R}_2$ & 5 & 0.121 & 0.202 & 0.164 & 0.153 & 0.360 \\
$\mathtt{R}_3\mathtt{R}_3$ & 7 & 0.045 & 0.229 & 0.504 & 0.184 & 0.039 \\
$\mathtt{R}_3\mathtt{R}_3\mathtt{R}_4$ & 3 & 0.320 & 0.368 & 0.215 & 0.097 & 0.000 \\
$\mathtt{R}_3\mathtt{R}_4$ & 3 & 0.159 & 0.183 & 0.369 & 0.203 & 0.086 \\
$\mathtt{R}_3\mathtt{R}_5$ & 3 & 0.333 & 0.217 & 0.159 & 0.068 & 0.222 \\
\hline
$\mathtt{R}_4$ & 10 & 0.147 & 0.295 & 0.252 & 0.180 & 0.126 \\
$\mathtt{R}_4\mathtt{R}_2$ & 4 & 0.163 & 0.301 & 0.198 & 0.188 & 0.150 \\
$\mathtt{R}_4\mathtt{R}_3$ & 4 & 0.137 & 0.226 & 0.226 & 0.266 & 0.146 \\
\hline
$\mathtt{R}_5$ & 8 & 0.118 & 0.200 & 0.302 & 0.236 & 0.145 \\
$\mathtt{R}_5\mathtt{R}_2$ & 3 & 0.143 & 0.182 & 0.521 & 0.057 & 0.098 \\
$\mathtt{R}_5\mathtt{R}_3$ & 3 & 0.186 & 0.183 & 0.195 & 0.344 & 0.091 \\
\hline
\end{tabular}
\end{minipage}}
\end{tabular}
\end{table}

\end{document}